\documentclass[twocolumn]{emulateapjdr}
\shorttitle{ {\it Kepler}'s Planetary Systems }
\shortauthors{Lissauer et al.}
\usepackage{natbib,ifthen,longtable}
\newcommand{\ikt}{{\it Kepler}}
\newcommand{\ik}{{\it Kepler~}}
\newcommand{\icarus}{{\it Icarus}}

\newcommand{\nsys}{961}
\newcommand{\npl}{1199}
\newcommand{\nall}{160171}

\begin{document} 
\slugcomment{accepted, to appear in ApJS, November 2011 issue}
\title{ Architecture and Dynamics of Kepler's\\ Candidate Multiple Transiting Planet Systems }

\author{Jack J. Lissauer\altaffilmark{1}, Darin Ragozzine\altaffilmark{2}, Daniel C. Fabrycky\altaffilmark{3,4}, Jason H. Steffen\altaffilmark{5},
Eric B. Ford\altaffilmark{6}, Jon M. Jenkins\altaffilmark{1,7}, Avi Shporer\altaffilmark{8,9}, Matthew J. Holman\altaffilmark{2}, Jason F. Rowe\altaffilmark{7}, Elisa V. Quintana\altaffilmark{7}, 
Natalie M. Batalha\altaffilmark{10}, William J. Borucki\altaffilmark{1}, Stephen T. Bryson\altaffilmark{1}, Douglas A. Caldwell\altaffilmark{7}, Joshua A. Carter\altaffilmark{2,4}, David Ciardi\altaffilmark{11}
Edward W. Dunham\altaffilmark{12}, Jonathan J. Fortney\altaffilmark{3}, Thomas N. Gautier, III\altaffilmark{13}, Steve B. Howell\altaffilmark{1}, 
David G. Koch\altaffilmark{1}, David W. 
Latham\altaffilmark{3}, Geoffrey W. Marcy\altaffilmark{14}, Robert C. Morehead\altaffilmark{6}, 
Dimitar Sasselov\altaffilmark{2}}

\email{ Jack.Lissauer@nasa.gov }
\altaffiltext{1}{NASA Ames Research Center, Moffett Field, CA 94035, USA}
\altaffiltext{2}{Harvard-Smithsonian Center for Astrophysics, 60 Garden Street, Cambridge, MA 02138, USA}
\altaffiltext{3}{Department of Astronomy \& Astrophysics, University of California, Santa Cruz, CA 95064, USA}
\altaffiltext{4}{Hubble Fellow}
\altaffiltext{5}{Fermilab Center for Particle Astrophysics, P.O. Box 500, MS 127, Batavia, IL 60510, USA}
\altaffiltext{6}{University of Florida, 211 Bryant Space Science Center, Gainesville, FL 32611, USA}
\altaffiltext{7}{SETI Institute/NASA Ames Research Center, Moffett Field, CA 94035, USA}
\altaffiltext{8}{Las Cumbres Observatory Global Telescope Network, 6740 Cortona Drive, Suite 102, Santa Barbara, CA 93117, USA}
\altaffiltext{9}{Department of Physics, Broida Hall, University of California, Santa Barbara, CA 93106, USA}
\altaffiltext{10}{Department of Physics and Astronomy, San Jose State University, San Jose, CA 95192, USA}
\altaffiltext{11}{Exoplanet Science Institute/Caltech, Pasadena, CA 91125, USA}
\altaffiltext{12}{Lowell Observatory, 1400 W. Mars Hill Road, Flagstaff, AZ 86001, USA}
\altaffiltext{13}{Jet Propulsion Laboratory, California Institute of Technology, Pasadena, CA 91109, USA}
\altaffiltext{14}{Astronomy Department, UC Berkeley, Berkeley, CA 94720, USA}

\begin{abstract}

About one-third of the $\sim$1200 transiting planet candidates detected in the first four months of \ik data are members of multiple candidate systems. 
There are 115 target stars with
two candidate transiting planets, 45 with three, 8 with four, and one each with five and six. We characterize the dynamical properties of these 
candidate multi-planet systems. The distribution of observed period ratios shows that the vast majority of candidate pairs are neither in nor near 
low-order mean motion resonances. Nonetheless, there are small but statistically significant excesses of candidate pairs both in resonance and spaced 
slightly too far apart to be in resonance, particularly near the 2:1 resonance. We find that virtually all candidate systems are stable, as tested by 
numerical integrations that assume a nominal mass-radius relationship. Several considerations strongly suggest that the vast majority of these 
multi-candidate systems are true planetary systems. Using the observed multiplicity frequencies, we find that a single population of planetary systems that matches the 
higher multiplicities underpredicts the number of singly-transiting systems. We provide constraints on the true multiplicity and mutual 
inclination distribution of the multi-candidate systems, revealing a population of systems with multiple super-Earth-size and Neptune-size planets with low to moderate mutual inclinations.  

\end{abstract}
\keywords{Celestial mechanics -- planetary systems; Planets and satellites: dynamical evolution and stability; Planetary systems; Planets and satellites: fundamental parameters; Planets and satellites: general; Facilities: Kepler Space Telescope}

\clearpage

\section{ Introduction }

\ik is a 0.95 m aperture space telescope that uses transit photometry to determine the frequency and characteristics of planets and 
planetary systems \citep{Borucki:2010, Koch:2010, Jenkins:2010, Caldwell:2010}. The focus of this NASA Discovery mission's design was to search for small planets in the habitable 
zone, but \emph{Kepler}'s ultra-precise, long-duration photometry is also ideal for detecting systems with multiple transiting planets. 
\cite{Borucki:2010a} presented 5 \ik targets with multiple transiting planet candidates, and these candidate planetary
systems were analyzed by \cite{Steffen:2010}. A system of three planets (Kepler-9=KOI-377, \citealt{Holman:2010a, Torres:2011}) was 
then reported, in which a near-resonant effect on the transit times allowed \ik data to confirm two of the planets and characterize the dynamics of 
the system. A compact system of six transiting planets \citep[Kepler-11,][]{Lissauer:2011}, five of which were confirmed by their 
mutual gravitational interactions exhibited through transit timing variations (TTVs), has also been reported.  These systems provide important data for understanding the dynamics, formation and evolution of planetary systems \citep{Koch:1996, 
Agol:2005, Holman:2005, Fabrycky:2009a, Holman:2010b, Ragozzine:2010}.

A catalog of 1235 candidate planets evident in the first four and a half months of \ik data is presented in \citep[][henceforth B11]{Borucki:2011}. Our 
analysis is based on the data presented by B11, apart from the following modifications: Five candidates identified as very likely false positives by 
\cite{Howard:2011}, Kepler Objects of Interest (KOIs) 1187.01, 1227.01, 1387.01, 1391.01 and 1465.01, are removed from the list.  The radius of KOI-1426.03 is reduced from 35 
R$_\oplus$ to 13 R$_\oplus$ (Section 4).  B11 reported that KOI-730.03's period was almost identical to that of  KOI-730.02, putting these two planets in the 1:1 resonance. Upon 
more careful lightcurve fits, we disfavor the co-orbital period of KOI-730.03 near 
10 days as an alias.  The period of KOI-730.03 is therefore increased to 19.72175 and its radius increased to 2.4 R$_\oplus$.  We do not 
consider the 18 candidates for which only a single transit was observed (and thus the period is poorly-constrained), nor the 13 remaining candidates 
with estimated radii $R_p > 22.4$ R$_\oplus$.  None of the removed candidates is in a multi-candidate system, but both of the candidates with adjusted 
parameters are.  We are left with a total of 1199 planetary candidates, 408 of which are in multiple candidate systems, around 961 target stars.

B11 also summarizes the observational data and follow-up programs 
being conducted to 
confirm and characterize planetary candidates.  Each object in this catalog is assigned a vetting flag (reliability score), with 1 signifying a confirmed or validated planet ($>\!\!98\%$ confidence), 2 being a strong candidate ($>\!\!80\%$ confidence), 3 being a somewhat weaker candidate ($>\!\!60\%$ confidence), and 4 signifying a candidate that has 
not received ground-based follow-up (this sample also has an expected reliability of $>\!\!60\%$). For most of the 
remainder of the paper, we refer to all of these objects as ``planets'', although 99\% remain unvalidated and thus 
should more properly be termed ``planet candidates.''   \cite{Ragozzine:2010}, \cite{Lissauer:2011}, and \cite{Latham:2011} point 
out that KOIs of target stars with multiple planet candidates are less likely to be caused by astrophysical false positives than are KOIs of stars 
with single candidates, although the possibility that one of the candidate signals is due to, e.g., a background eclipsing binary, can be non-negligible.

\begin{figure*}
\includegraphics[width=0.45\textwidth,angle=90]{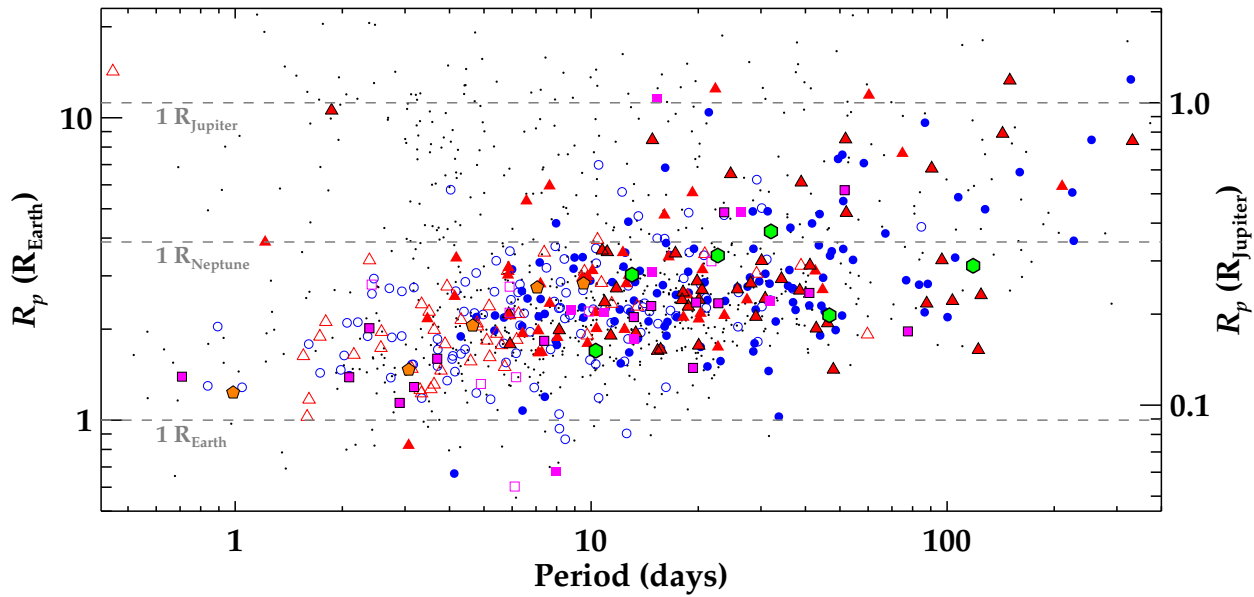}
\caption{Planet period vs. radius for all \npl ~planetary candidates from B11 that we consider herein.  Those planets that are the only candidate for their given star are represented by black dots, those in two-planet systems as blue circles (open for the inner planets, filled for the outer ones), those in three-planet systems as red triangles (open for the inner planets, filled for the middle ones, filled with black borders for the outer ones), those in four planet systems as  purple squares (inner and outer members filled with black borders, second members open, third filled), the five candidates of KOI-500 as orange pentagons and the six planets orbiting KOI-157 (Kepler-11) as green hexagons. It is immediately apparent that there is a paucity of giant planets in multi-planet systems; this difference in the size distributions is quantified and discussed by \cite{Latham:2011}.  The upward slope in the lower envelope of these points is caused by the low SNR of small transiting planets with long orbital periods (for which few transits have thus far been observed).  Figure provided by Samuel Quinn.}
\label{fig:perrad}
\end{figure*}

The apparently single systems reported by B11 are likely to include many systems that possess additional planets that have thus far escaped detection by being too small, 
too long-period, or too highly inclined with respect to the line of sight.  \cite{Ford:2011} 
discuss candidates exhibiting transit timing variations (TTVs) that may be caused by these non-transiting planets, but so far no non-transiting planets have been identified.
\cite{Latham:2011} compare various orbital and physical properties of 
planets in single candidate systems to those of planets in multi-candidate systems.  We examine herein the dynamical aspects of {\it Kepler}'s multi-planet systems.

Multi-transiting systems provide numerous insights that are difficult or impossible to gain from 
single-transiting systems \citep{Ragozzine:2010}. These systems harness the power of radius measurements from transit photometry in combination with the 
illuminating properties of multi-planet orbital architecture. Observable interactions between planets in these systems, seen in TTVs, are much easier to characterize, and our knowledge of both stellar and planetary parameters is improved. The 
true mutual inclinations between planets in multi-transiting systems, while not directly observable from the light curves themselves, are much 
easier to obtain than in other systems, both individually and statistically \citep[see Section \ref{sec:coplanar} and][]{Lissauer:2011}. The distributions of physical and orbital parameters of 
planets in these systems are invaluable for comparative planetology \citep[e.g.,][]{Havel:2011} and provide important insights into the processes of planet formation and evolution.  It is thus very exciting that B11 report 115 doubly, 45 triply, 8 quadruply, 1 quintuply and 1 sextuply transiting system. 

\tabletypesize{\tiny}
\def\arraystretch{0.9}

\begin{deluxetable}{ccccc}
\tabletypesize{\scriptsize}
\tablewidth{0pc}
\tablecaption{ Characteristics of Systems with Two Transiting Planets \label{tab:two} }
\tablehead{
\colhead{KOI \#} & \colhead{$R_{p,1}$ (R$_\oplus$)} & \colhead{$R_{p,2}$ (R$_\oplus$) } & \colhead{$P_2/P_1$} & \colhead{$\Delta$} }
\startdata
\hline
72 &   1.30 &   2.29 &    54.083733 &  90.8\\
82 &   3.70 &   6.84 &     1.565732 &   7.6\\
89 &   4.36 &   5.46 &     1.269541 &   4.8\\
112 &   1.71 &   3.68 &    13.770977 &  55.7\\
115 &   3.37 &   2.20 &     1.316598 &   7.4\\
116 &   4.73 &   4.81 &     3.230779 &  21.6\\
123 &   2.26 &   2.50 &     3.274199 &  34.1\\
124 &   2.33 &   2.83 &     2.499341 &  25.5\\
139 &   1.18 &   5.66 &    67.267271 &  54.4\\
150 &   3.41 &   3.69 &     3.398061 &  26.3\\
153 &   3.07 &   3.17 &     1.877382 &  14.2\\
209 &   4.86 &   7.55 &     2.702205 &  14.8\\
220 &   2.62 &   0.67 &     1.703136 &  17.3\\
222 &   2.06 &   1.68 &     2.026806 &  20.5\\
223 &   2.75 &   2.40 &    12.906148 &  55.8\\
232 &   1.55 &   3.58 &     2.161937 &  20.1\\
244 &   2.65 &   4.53 &     2.038993 &  15.5\\
260 &   1.19 &   2.19 &     9.554264 &  70.3\\
270 &   0.90 &   1.03 &     2.676454 &  49.1\\
271 &   1.82 &   1.99 &     1.654454 &  17.5\\
279 &   2.09 &   4.90 &     1.846201 &  13.4\\
282 &   0.86 &   2.77 &     3.252652 &  36.0\\
291 &   1.05 &   1.45 &     3.876548 &  57.0\\
313 &   2.17 &   3.10 &     2.220841 &  21.0\\
314 &   1.95 &   1.57 &     1.675518 &  14.8\\
339 &   1.47 &   1.08 &     3.240242 &  50.6\\
341 &   2.26 &   3.32 &     1.525757 &  11.0\\
343 &   1.64 &   2.19 &     2.352438 &  28.6\\
386 &   3.40 &   2.90 &     2.462733 &  21.7\\
401 &   6.24 &   6.61 &     5.480100 &  23.3\\
416 &   2.95 &   2.82 &     4.847001 &  35.8\\
431 &   3.59 &   3.50 &     2.485534 &  19.6\\
433 &   5.78 &  13.40 &    81.440719 &  28.7\\
440 &   2.24 &   2.80 &     3.198296 &  30.0\\
442 &   1.43 &   1.86 &     7.815905 &  67.9\\
446 &   2.31 &   1.69 &     1.708833 &  15.5\\
448 &   2.33 &   3.79 &     4.301984 &  28.2\\
456 &   1.66 &   3.12 &     3.179075 &  31.6\\
459 &   1.28 &   3.69 &     2.810266 &  26.6\\
464 &   2.66 &   7.08 &    10.908382 &  33.2\\
474 &   2.34 &   2.32 &     2.648401 &  28.7\\
475 &   2.36 &   2.63 &     1.871903 &  17.0\\
490 &   2.29 &   2.25 &     1.685884 &  15.1\\
497 &   1.72 &   2.49 &     2.981091 &  33.8\\
508 &   3.76 &   3.52 &     2.101382 &  16.3\\
509 &   2.67 &   2.86 &     2.750973 &  25.6\\
510 &   2.67 &   2.67 &     2.172874 &  20.6\\
518 &   2.37 &   1.91 &     3.146856 &  32.3\\
523 &   2.72 &   7.31 &     1.340780 &   4.9\\
534 &   1.38 &   2.05 &     2.339334 &  28.7\\
543 &   1.53 &   1.89 &     1.371064 &  11.1\\
551 &   1.79 &   2.12 &     2.045843 &  23.5\\
555 &   1.51 &   2.27 &    23.366026 &  77.3\\
564 &   2.35 &   4.99 &     6.073030 &  34.0\\
573 &   2.10 &   3.15 &     2.908328 &  28.3\\
584 &   1.58 &   1.51 &     2.138058 &  28.5\\
590 &   2.10 &   2.22 &     4.451198 &  44.3\\
597 &   1.40 &   2.60 &     8.272799 &  59.7\\
612 &   3.51 &   3.62 &     2.286744 &  18.1\\
638 &   4.76 &   4.15 &     2.838628 &  19.5\\
645 &   2.64 &   2.48 &     2.796998 &  28.9\\
657 &   1.59 &   1.90 &     4.001270 &  42.5\\
658 &   1.53 &   2.22 &     1.698143 &  18.1\\
663 &   1.88 &   1.75 &     7.369380 &  52.0\\
672 &   3.98 &   4.48 &     2.595003 &  18.8\\
676 &   2.94 &   4.48 &     3.249810 &  20.4\\
691 &   1.29 &   2.88 &     1.828544 &  18.8\\
693 &   1.72 &   1.80 &     1.837696 &  22.2\\
700 &   1.93 &   3.05 &     3.297027 &  32.2\\
708 &   1.71 &   2.19 &     2.262657 &  27.3\\
736 &   1.96 &   2.64 &     2.789104 &  24.7\\
738 &   3.30 &   2.86 &     1.285871 &   6.2\\
749 &   1.35 &   1.98 &     1.357411 &  11.0\\
752 &   2.69 &   3.39 &     5.734874 &  39.3\\
775 &   2.47 &   2.11 &     2.079976 &  17.9\\
787 &   2.87 &   2.19 &     1.284008 &   7.0\\
800 &   2.73 &   2.52 &     2.659942 &  26.3\\
837 &   1.45 &   1.83 &     1.919049 &  22.3\\
841 &   4.00 &   4.91 &     2.042803 &  13.0\\
842 &   2.78 &   3.14 &     2.835637 &  22.9\\
853 &   2.87 &   2.12 &     1.767103 &  15.2\\
869 &   3.19 &   4.33 &     4.845276 &  30.2\\
870 &   3.62 &   3.45 &     1.519922 &   8.9\\
877 &   2.50 &   2.32 &     2.021803 &  17.0\\
881 &   2.54 &   3.92 &    10.793324 &  43.8\\
896 &   2.81 &   3.85 &     2.574418 &  20.6\\
904 &   2.11 &   2.96 &    12.635882 &  50.2\\
936 &   2.04 &   3.46 &    10.601824 &  40.6\\
938 &   1.28 &   2.89 &     9.512348 &  57.5\\
945 &   2.04 &   2.64 &     1.575035 &  13.6\\
954 &   2.35 &   2.45 &     4.550137 &  40.9\\
1015 &   1.63 &   2.36 &     2.305817 &  27.5\\
1060 &   1.23 &   1.54 &     2.545144 &  39.7\\
1089 &   5.69 &   9.64 &     7.094116 &  22.6\\
1102 &   0.94 &   3.22 &     1.513919 &  12.3\\
1113 &   2.65 &   2.81 &     3.217321 &  30.7\\
1151 &   1.17 &   1.20 &     1.420286 &  16.6\\
1163 &   1.90 &   1.77 &     2.729492 &  33.8\\
1198 &   1.53 &   2.04 &     1.561840 &  16.0\\
1203 &   2.44 &   2.51 &     2.256556 &  23.0\\
1215 &   2.21 &   2.12 &     1.905318 &  20.8\\
1221 &   5.03 &   5.31 &     1.694236 &  10.0\\
1236 &   1.67 &   2.78 &     5.807534 &  49.5\\
1241 &   6.99 &  10.43 &     2.039957 &   9.3\\
1278 &   1.57 &   2.39 &     3.575571 &  40.0\\
1301 &   1.86 &   2.32 &     2.954666 &  32.3\\
1307 &   2.81 &   2.96 &     2.204782 &  20.1\\
1360 &   2.29 &   2.73 &     2.520266 &  24.0\\
1364 &   2.74 &   2.86 &     2.952861 &  27.1\\
1396 &   1.86 &   2.53 &     1.790301 &  17.8\\
1475 &   1.79 &   2.18 &     5.910858 &  43.3\\
1486 &   2.38 &   8.45 &     8.433622 &  27.9\\
1589 &   2.23 &   2.28 &     1.476360 &  12.0\\
1590 &   1.90 &   2.75 &    10.943201 &  55.8\\
1596 &   2.26 &   3.45 &    17.785655 &  54.0
\enddata
\end{deluxetable}

\tabletypesize{\scriptsize}
\def\arraystretch{1.000}

We begin by summarizing the characteristics of \ikt's multi-planet systems in Section 2.  The reliability of the data set is discussed in Section 3.  
Sections 4 -- 6 present our primary statistical results. Section 4 presents an analysis of the long-term dynamical stability of candidate planetary 
systems for nominal estimates of planetary masses and unobserved orbital parameters.  Evidence of orbital commensurabilities (mean motion resonances) among planet candidates 
is analyzed in Section 5; this section also includes specific comments on some of the most interesting cases in the reported systems. The characteristic distributions of numbers of planets per system and mutual inclinations of planetary orbits are discussed 
in Section 6.  We compare the Kepler-11 planetary system \citep{Lissauer:2011} to other \ik candidates in Section 7.  We conclude by discussing the 
implications of our results and summarize the most important points presented herein.

\section{Characteristics of {\it Kepler} Multi-Planet Systems}  \label{sec:char}

The light curves derived from \ik data can be used to measure the radii and orbital periods of planets.  Periods are measured to high accuracy. The ratios of planetary radii to stellar radii are also well measured, except for low signal to noise transits 
(e.g., planets much smaller than their 
stars, those for which few transits are observed, and planets orbiting faint or variable stars).  The primary source of errors in planetary radii are 
probably errors in estimated stellar radii, which are calculated using photometry from the \emph{Kepler} Input Catalog (B11).  In some cases, the 
planet's radius is underestimated because a significant fraction of the flux in the target aperture is provided by a star other than the one being 
transited by the planet.  This extra light can often be determined and corrected for using data from the spacecraft \citep{Bryson:2010} and/or from the 
ground, although dilution by faint physically-bound stars is difficult to detect~\citep{Torres:2011}.  Spectroscopic studies can, and in the future 
will, be used to estimate the stellar radii, thereby allowing for improved estimates of the planetary radii; this procedure will also provide more 
accurate estimates of the stellar masses.

\begin{deluxetable*}{cccccccc}
\tabletypesize{\scriptsize}
\tablewidth{0pc}
\tablecaption{ Characteristics of Systems with Three Transiting Planets \label{tab:three} }
\tablehead{
\colhead{KOI \#} & \colhead{$R_{p,1}$ (R$_\oplus$)} & \colhead{$R_{p,2}$ (R$_\oplus$)} &  \colhead{$P_2/P_1$} & \colhead{$\Delta_{1,2}$} & \colhead{$R_{p,3}$ (R$_\oplus$) } & \colhead{$P_3/P_2$} & \colhead{$\Delta_{2,3}$}  }
\startdata
\hline
85 &   1.67 &   3.24 &     2.719393 &  28.2 &   2.01 &     1.387583 &   9.2\\
94 &   4.01 &  12.64 &     2.143302 &   9.0 &   6.89 &     4.052306 &  14.9\\
111 &   2.46 &   2.25 &     2.071194 &  21.5 &   2.51 &     4.373353 &  40.8\\
137 &   2.32 &   6.04 &     2.180367 &  14.4 &   8.56 &     1.944495 &   8.9\\
148 &   2.12 &   2.97 &     2.024651 &  18.7 &   2.04 &     4.434225 &  37.6\\
152 &   2.41 &   2.54 &     2.032345 &  20.4 &   4.92 &     1.900850 &  13.5\\
156 &   1.64 &   1.90 &     1.549841 &  14.0 &   2.77 &     1.464437 &  10.0\\
168 &   1.86 &   2.00 &     1.395701 &  11.9 &   3.69 &     1.511764 &  10.9\\
248 &   1.99 &   2.86 &     2.795806 &  23.1 &   2.49 &     1.515099 &   9.1\\
250 &   1.28 &   3.63 &     3.465802 &  25.5 &   3.61 &     1.404620 &   6.1\\
284 &   1.87 &   1.96 &     1.038334 &   1.3 &   2.53 &     2.807616 &  31.5\\
351 &   1.95 &   6.02 &     3.522872 &  23.1 &   8.50 &     1.575869 &   6.2\\
377 &   1.04 &   5.73 &    12.089890 &  40.6 &   6.20 &     2.020508 &  10.7\\
398 &   1.92 &   3.49 &     2.417105 &  21.8 &   8.61 &    12.403143 &  29.2\\
408 &   3.63 &   2.87 &     1.701564 &  12.5 &   2.56 &     2.454333 &  23.6\\
481 &   1.65 &   2.47 &     4.922978 &  45.4 &   2.97 &     4.478249 &  36.0\\
520 &   1.96 &   3.06 &     2.348524 &  22.6 &   2.75 &     2.018139 &  17.2\\
528 &   3.14 &   3.19 &     2.146104 &  18.2 &   3.43 &     4.703530 &  33.6\\
567 &   2.89 &   2.29 &     1.899657 &  17.5 &   2.23 &     1.429507 &  10.9\\
571 &   1.81 &   1.70 &     1.869767 &  17.7 &   1.97 &     1.836067 &  16.7\\
623 &   1.80 &   2.04 &     1.848406 &  21.8 &   1.74 &     1.514823 &  15.0\\
665 &   1.19 &   0.84 &     1.905525 &  33.9 &   2.28 &     1.910436 &  23.9\\
701 &   1.52 &   2.22 &     3.178366 &  35.2 &   1.73 &     6.737918 &  52.2\\
711 &   1.33 &   2.74 &    12.349935 &  64.8 &   2.63 &     2.785831 &  26.8\\
718 &   1.59 &   1.77 &     4.953555 &  54.9 &   1.49 &     2.108971 &  27.9\\
723 &   2.81 &   3.17 &     2.562570 &  22.2 &   2.87 &     2.783488 &  23.9\\
733 &   1.50 &   2.25 &     1.891175 &  20.2 &   1.93 &     1.915474 &  19.3\\
756 &   1.76 &   2.60 &     1.610874 &  14.8 &   3.65 &     2.683265 &  23.3\\
757 &   2.28 &   4.84 &     2.569790 &  18.5 &   3.29 &     2.563540 &  17.3\\
806 &   3.10 &  12.05 &     2.068502 &   8.3 &   8.99 &     2.373352 &   8.7\\
812 &   2.45 &   2.19 &     6.005814 &  39.0 &   2.12 &     2.302249 &  20.8\\
829 &   1.93 &   2.63 &     1.912284 &  19.4 &   2.72 &     2.067646 &  19.5\\
864 &   2.17 &   1.82 &     2.265275 &  25.5 &   1.79 &     2.052797 &  24.3\\
884 &   1.24 &   3.00 &     2.829393 &  27.8 &   2.74 &     2.169267 &  18.1\\
898 &   2.41 &   3.03 &     1.889897 &  13.8 &   2.54 &     2.056092 &  15.3\\
899 &   1.28 &   1.69 &     2.151439 &  22.7 &   1.72 &     2.160301 &  20.8\\
907 &   2.07 &   3.52 &     3.446956 &  30.7 &   3.41 &     1.824674 &  13.7\\
921 &   1.48 &   2.30 &     2.717125 &  31.1 &   2.68 &     1.762252 &  15.2\\
934 &   3.16 &   2.02 &     2.130198 &  20.2 &   2.41 &     1.510392 &  12.6\\
935 &   3.59 &   3.16 &     2.043775 &  16.5 &   2.47 &     2.055839 &  18.8\\
941 &   3.43 &   5.39 &     2.762270 &  18.3 &   6.61 &     3.747567 &  19.1\\
961 &  14.44 &   3.94 &     2.677706 &   8.5 &  10.72 &     1.536625 &   4.6\\
1306 &   2.14 &   2.20 &     1.930557 &  20.5 &   1.81 &     1.705578 &  17.6\\
1422 &   2.02 &   3.06 &     1.613089 &  10.5 &   2.92 &     3.398094 &  23.4\\
1426 &   3.29 &   7.74 &     1.926989 &  10.3 &  12.30 &     2.002740 &   7.5
\enddata
\end{deluxetable*}

We list the key observed properties of \ik two-planet systems in Table~\ref{tab:two}, three-planet systems in Table~\ref{tab:three}, four-planet 
systems in Table~\ref{tab:four}, and those systems with more than four planets in Table~\ref{tab:fiveplus}. Planet indices given in these tables 
signify orbit order, with 1 being the planet with the shortest period.  These do not always correspond to the post-decimal point portion (.01, .02, 
etc.) of the KOI number designation of these candidates given in B11, for which the numbers signify the order in which the candidates were identified. 
These tables are laid out differently from one another because of the difference in parameters that are important for systems with differing numbers of 
planets.  In addition to directly observed properties, these tables contain results of the dynamical analyses discussed below.

\begin{deluxetable*}{ccccccccccc}
\tabletypesize{\scriptsize}
\tablewidth{0pc}
\tablecaption{ Characteristics of Systems with Four Transiting Planets \label{tab:four} }
\tablehead{
\colhead{KOI \#} & \colhead{$R_{p,1}$ (R$_\oplus$)} & \colhead{$R_{p,2}$ (R$_\oplus$)} &  \colhead{$P_2/P_1$} & \colhead{$\Delta_{1,2}$} & \colhead{$R_{p,3}$ (R$_\oplus$) } & \colhead{$P_3/P_2$} & \colhead{$\Delta_{2,3}$} & \colhead{$R_{p,4}$ (R$_\oplus$) } & \colhead{$P_4/P_3$} & \colhead{$\Delta_{3,4}$}  }
\startdata
\hline
70 &   1.60 &   0.60 &     1.649977 &  22.6 &   2.27 &     1.779783 &  20.8 &   1.97 &     7.150242 &  53.5\\
117 &   1.29 &   1.32 &     1.541373 &  19.4 &   0.68 &     1.623476 &  25.1 &   2.39 &     1.853594 &  22.3\\
191 &   1.39 &   2.81 &     3.412824 &  35.8 &  11.56 &     6.350795 &  20.2 &   1.49 &     1.258207 &   2.9\\
707 &   2.19 &   3.36 &     1.652824 &  13.4 &   2.48 &     1.459577 &   9.8 &   2.64 &     1.290927 &   7.3\\
730 &   1.83 &   2.29 &     1.333812 &   9.4 &   3.09 &     1.500997 &  11.0 &   2.63 &     1.333948 &   7.5\\
834 &   1.39 &   1.39 &     2.943873 &  44.2 &   1.85 &     2.149832 &  28.5 &   4.87 &     1.787503 &  12.5\\
880 &   2.01 &   2.76 &     2.476866 &  25.6 &   4.90 &     4.480164 &  28.7 &   5.76 &     1.948715 &  11.1\\
952 &   1.14 &   2.27 &     2.037706 &  19.5 &   2.32 &     1.483153 &   9.2 &   2.44 &     2.602734 &  21.3
\enddata
\end{deluxetable*}

\begin{deluxetable*}{cccc} 
\tabletypesize{\scriptsize}
\tablewidth{0pc}
\tablecaption{ Characteristics of Systems with Five or Six Transiting Planets \label{tab:fiveplus} } 
\tablehead{
\colhead{property} & \colhead{KOI-500} & \colhead{KOI-157 (nominal Kepler-11)} & \colhead{Kepler-11 \citep{Lissauer:2011}} } 
\startdata 
\hline
 {$R_{p,1}$ (R$_\oplus$)} & 1.23 &      1.70 &  1.97$\pm$0.19\\
{$R_{p,2}$ (R$_\oplus$)} & 1.47 &  3.03 &  3.15$\pm$0.30\\
 {$P_2/P_1$} & 3.113327 &   1.264079 & 1.26410 \\
  {$\Delta_{12}$} & 40.5 &     6.8 &  7.0 \\
 {$R_{p,3}$ (R$_\oplus$) } &  2.06 & 3.50 &  3.43$\pm$0.32 \\
{$P_3/P_2$} &   1.512077 & 1.741808 & 1.74182 \\
{$\Delta_{23}$} & 12.7 &   13.1 &  15.9 \\ 
 {$R_{p,4}$ (R$_\oplus$) } & 2.75 & 4.21 &  4.52$\pm$0.43 \\
 {$P_4/P_3$} &  1.518394 &   1.410300 &    1.41031 \\
 {$\Delta_{34}$}  &     10.4 & 7.3 & 10.9 \\
 {$R_{p,5}$ (R$_\oplus$) } &  2.83 &  2.22  &  2.61$\pm$0.25 \\
 {$P_5/P_4$} &  1.349929 &    1.459232 &  1.45921 \\
   {$\Delta_{45}$}  &    6.8 &       8.8 &  13.3 \\ 
    {$R_{p,6}$ (R$_\oplus$) } & \nodata  &  3.24  &  3.66$\pm$0.35 \\
  {$P_6/P_5$} & \nodata &    2.535456 & 2.53547 \\
 {$\Delta_{56}$} & \nodata &   24.1  & \nodata
\enddata
\end{deluxetable*}

\begin{figure}
\includegraphics[width=0.45\textwidth]{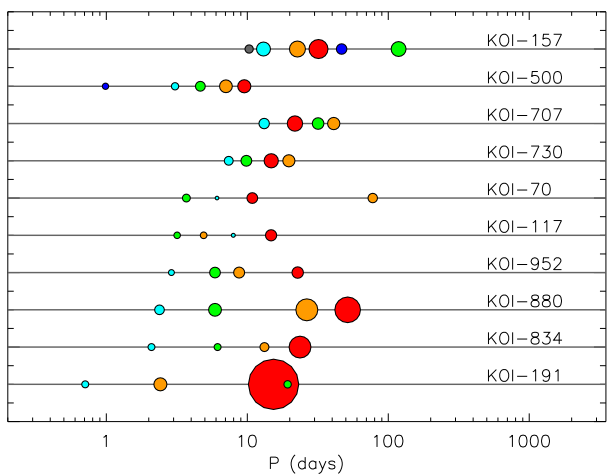}
\caption{ 
Gallery of candidate planetary systems of four to six planets.  Each horizontal line represents a separate system, as labeled.  They are sorted by, 
first, the number of candidates, and second, the innermost planet's orbital period.  The size of each dot is proportional to the size of the planet that it represents, and 
in each system the size orderings move from hot (large) to cool (small) colors (red for the largest planet within its system, then gold, green, aqua, 
and, if additional planets are present, navy and gray). There is a clear trend for smaller planets
to be interior to larger planets, but this is due to the greater detectability of small planets at shorter orbital period.}
\label{fig:fourplus}
\end{figure}

\begin{figure}
\includegraphics[width=0.45\textwidth]{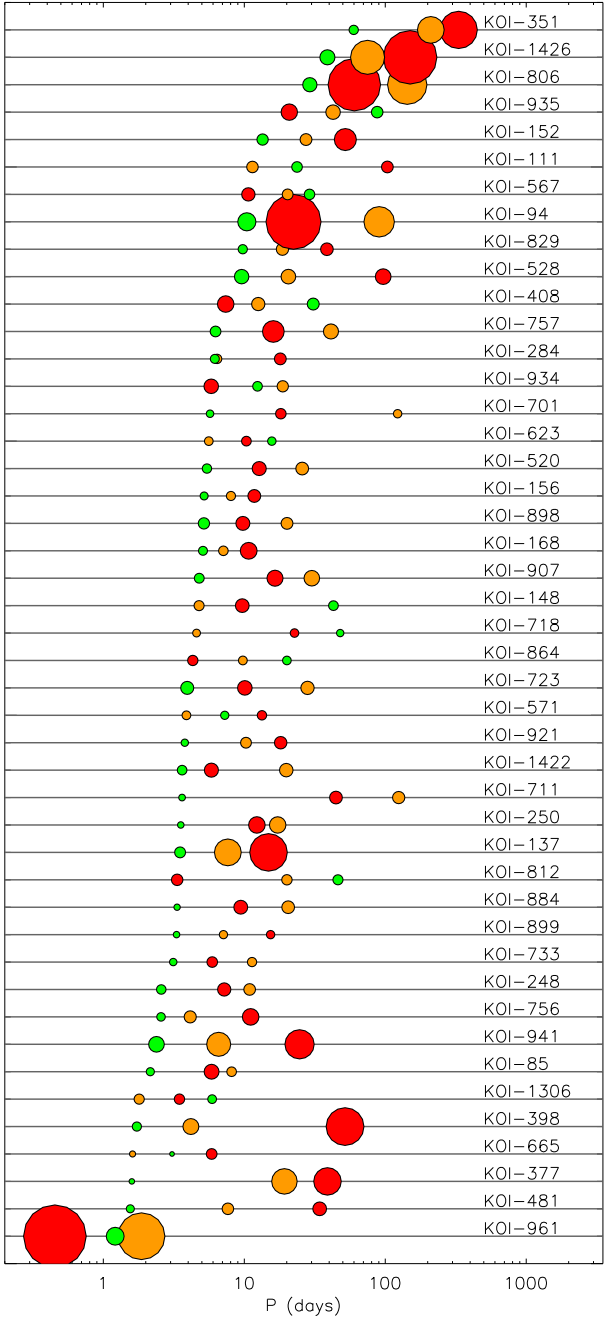}
\caption{ Three planet candidate systems; same format as Figure~\ref{fig:fourplus}.}
\label{fig:triples}
\end{figure}

\begin{figure*}
\includegraphics[width=0.9\textwidth]{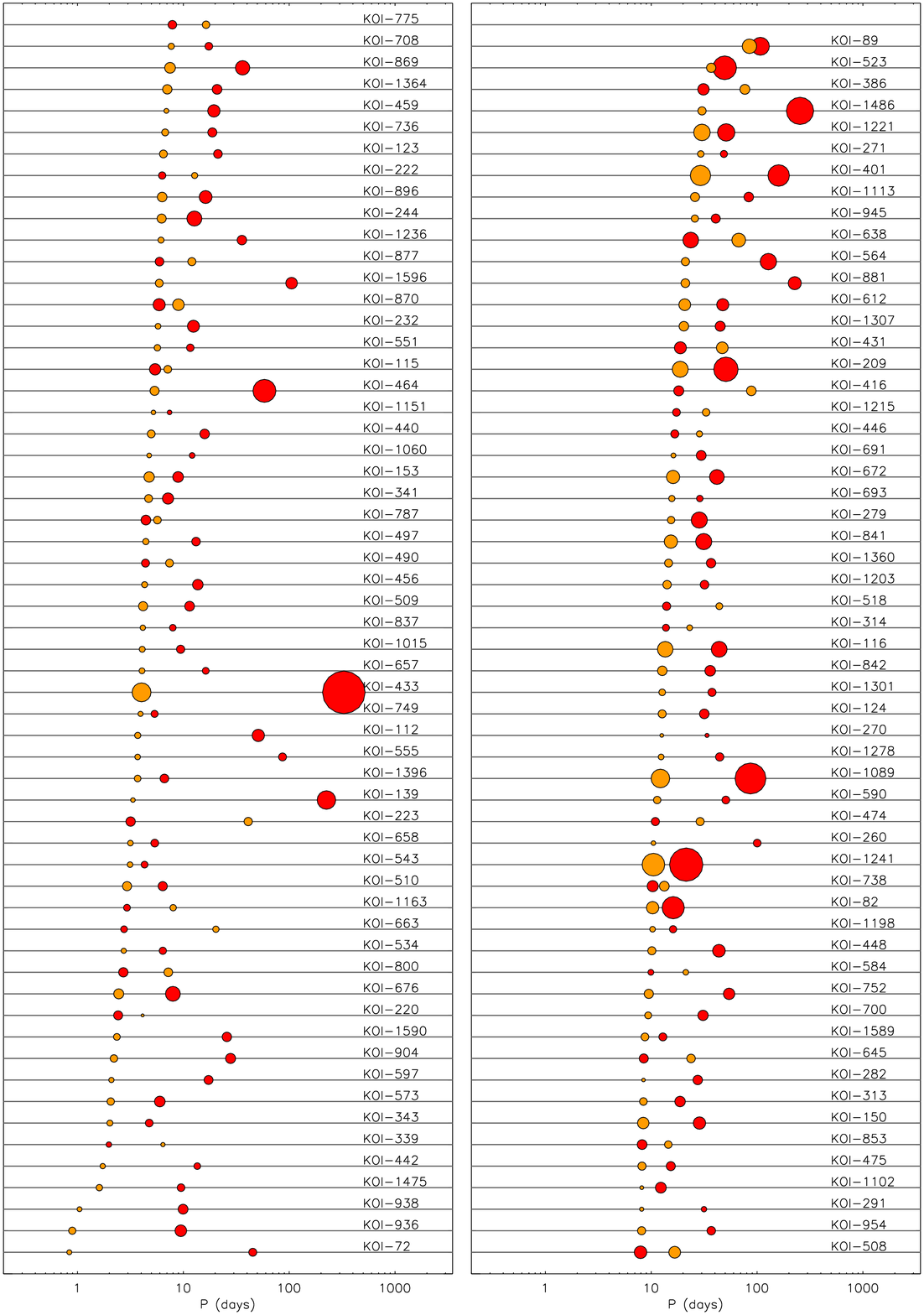}
\caption{ Two planet candidate systems; same format as Figure~\ref{fig:fourplus}.}
\label{fig:doubles}
\end{figure*}

\begin{figure}
\includegraphics[width=0.5\textwidth]{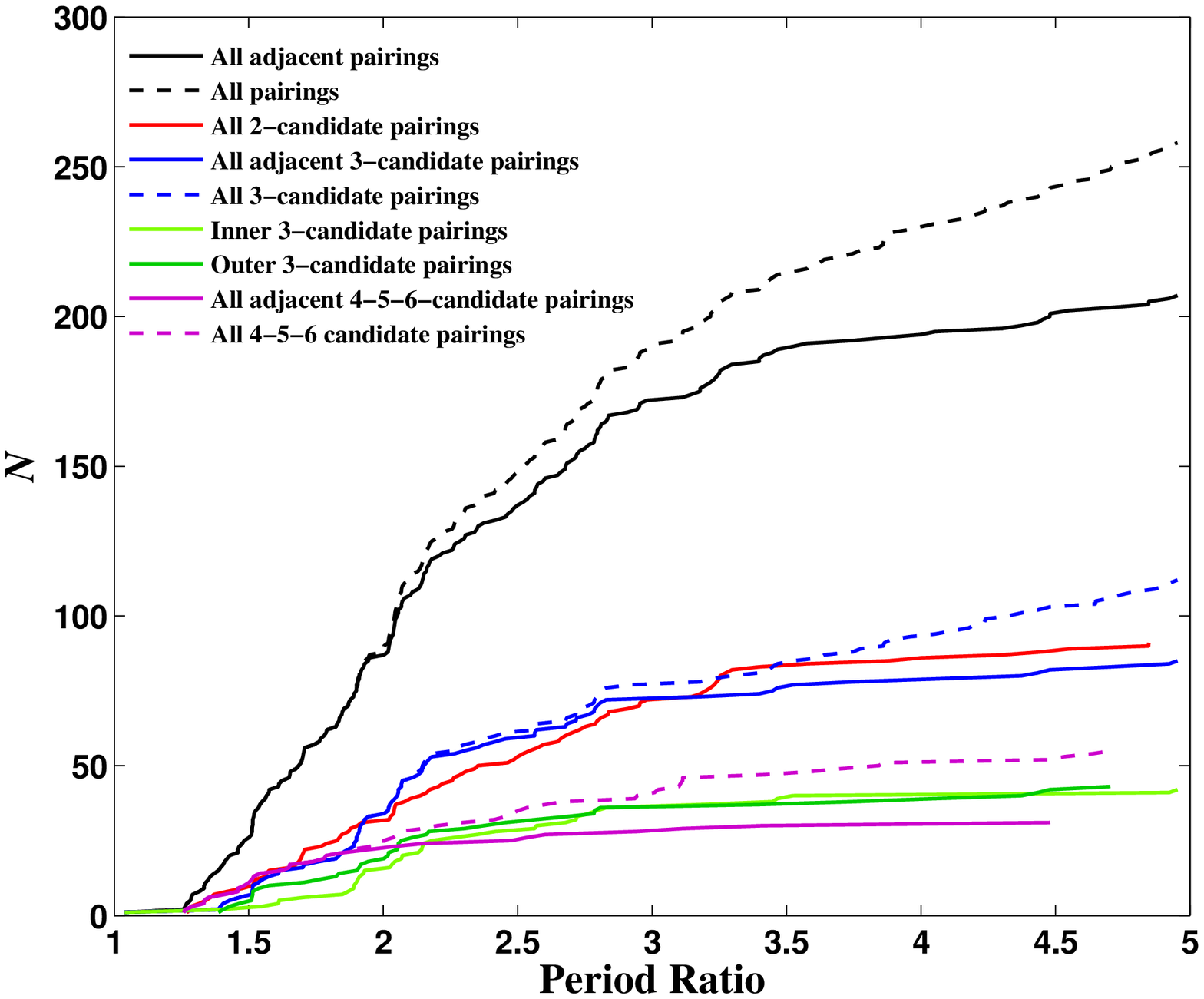}

\caption{Cumulative number, $N$, of pairs of \ik planets orbiting the same star with period ratio, ${\cal P}$ ($\equiv P_o/P_i$), less than the value 
specified.  Black dashed 
curve shows all pairings, solid black curve shows pairings of neighboring planets; solid red curve shows all pairings in two candidate systems; dashed 
blue curve shows all pairings in three-planet systems, solid 
blue curve shows all adjacent pairings in three-planet systems, solid light green curve shows inner pair of planets in three-planet systems, the solid dark 
green curve shows outer pair of planets in three-planet systems, the dashed pink curve shows all pairings in the 4, 5 and 6 planet systems, solid 
pink curve shows neighboring pairings in the 4, 5 and 6 planet system.  Sixty-five pairs of planets in the same system, including 31 adjacent 
pairs, have ${\cal P} > 5$ and are not represented on this plot.} \label{fig:cumprat} \end{figure}

\begin{figure}
\includegraphics[width=0.5\textwidth]{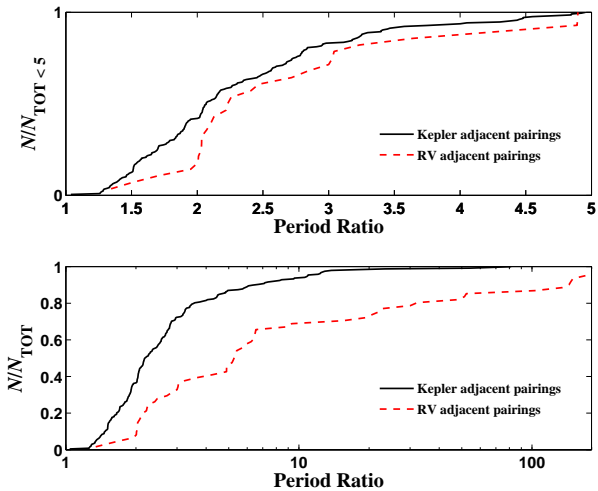}
\caption{
Cumulative fraction of neighboring planet pairs for \ik candidate multi-planet systems with period ratio less than the value 
specified (solid curve).  The cumulative fraction for neighboring pairs in multi-planet systems detected via radial velocity is also 
shown (dashed curve), and includes data from exoplanets.org as of 29 January 2011.
(a) Linear horizontal axis, as in Figure~\ref{fig:cumprat}.  The data are normalized for the number of adjacent pairs with $\cal P$ $< 5$, which equals 207 for the \ik candidates and 28 for RV planets. 
(b) Logarithmic horizontal axis.  All 238 \ik pairs are shown; the three RV pairs with $\cal P >$ 200 are omitted from the plot, but used for the 
calculation of $N_{tot}$, which is equal to 61.
}
\label{fig:rvcumprat}
\end{figure}

\begin{figure}
\includegraphics[width=0.5\textwidth]{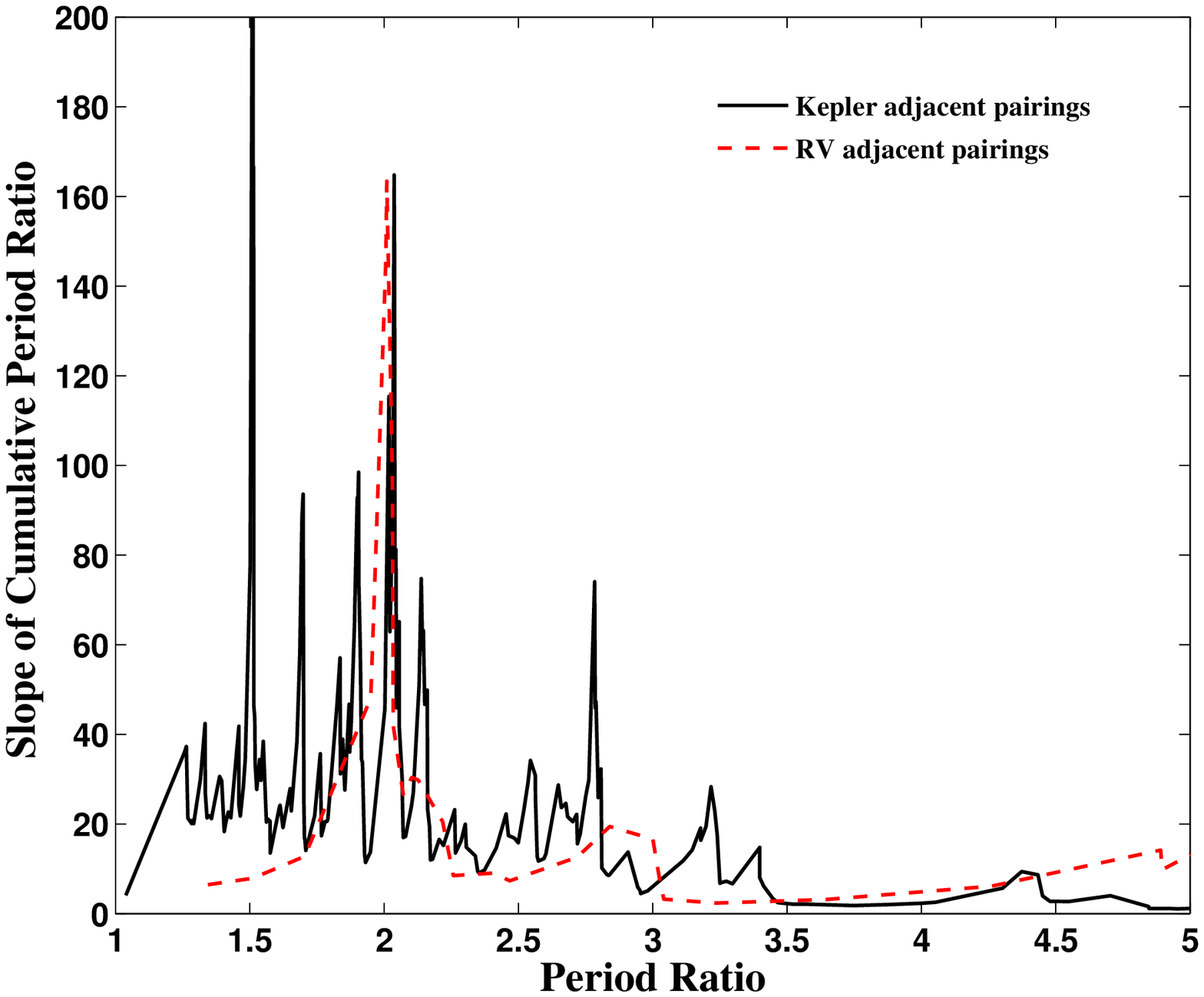}
\caption{Slope of the cumulative fraction of \ik neighboring planet pairs (solid black curve) and multi-planet systems detected via radial velocity 
(dashed red curve) with period ratio less than the value specified. The slope for the \ik curve was computed by taking the difference in period ratio 
between points with $N$ differing by 4 and dividing this difference by 4. The slope for the RV curve was computed by taking the difference in period ratio between 
points with $N$ differing by 3 and dividing by 3, and then normalizing by multiplying the value by the ratio of the number of \ik pairings to the 
number of radial velocity pairings (3.9). The spikes in both curves near $\cal P$ = 2 and the \ik curve near $\cal P$ = 1.5 show excess planets piling up near period commensurabilities.  Two points on the \ik curve near period ratio 1.5 lie above the plot, with the highest value being 299.84; this sharp peak is produced by the excess of observed planet pairs in or very near the 3:2 mean motion resonance.}
\label{fig:deriv}
\end{figure}

We plot the period and radius of each active \ik planetary candidate observed during the first four and one-half months of spacecraft operations, with 
special emphasis on multi-planet systems, in Figure~\ref{fig:perrad}.  We present galleries of multi-planet candidates, representing the planetary sizes 
and periods, in Figures~\ref{fig:fourplus}--\ref{fig:doubles}.  The ratio(s) of orbital periods is a key factor in planetary system dynamics; the 
cumulative distributions of these period ratios for various classes of planetary pairings are plotted in Figure~\ref{fig:cumprat}.  
Figure~\ref{fig:rvcumprat} compares the cumulative distribution of the period ratio of neighboring pairs of \ik planet candidates with the comparable 
distribution for radial velocity planets. The derivatives of these distributions with respect to period ratio, which show spikes near common period 
ratios, are displayed in Figure~\ref{fig:deriv}.

The probability that a \ik target star hosts at least one detected transiting planet is {\nsys}/{\nall} $\approx$ 0.006, while the probability that a star 
with 
one detected transiting planet hosts at least one more is much higher, 170/\nsys ~$\approx$ 0.177. To determine whether or not the observed 
transiting 
planets are randomly distributed among host stars, we compare the observed distribution of planets per system to three random populations.  In the 
first of these synthetic populations, each of \npl~planets is randomly assigned to one of \nall ~stars. The number of stars containing $j$ planets in 
this random population is equivalent to a Poisson distribution with a mean of \npl/\nall ~$\approx$ 0.0075, and is denoted as ``Poisson 0''. For the 
second synthetic population, ``Poisson 1'', we again use a Poisson distribution, also forced to match the observed total number of \ik planets, \npl, 
but in this case the second requirement is to match the total number of planetary systems (stars with at least 1 observed planet), \nsys, rather than 
the total number of stars, \nall.  Results are shown in Table~\ref{tab:Poisson}, which also includes comparisons to a third random distribution 
discussed below and to the population of planets detected by radial velocity surveys normalized to the same total number of planets as the \ik candidates 
considered 
here. There are sharp differences between \ik observations and Poisson 0: The number of observed \ik multi-transiting systems greatly exceeds the 
random distribution, showing that transiting planets tend to come in planetary systems, i.e., once a planet is detected to transit a given star, that 
star is much more likely to be orbited by an additional transiting planet than would be the case if transiting planets were randomly distributed among targets.  Poisson 1, the random 
distribution that fits the total number of planets and the total number of stars with planets, underpredicts the numbers of systems with three or more 
planets, suggesting that there is more than one population of planetary systems.  To carry the study further, we used a third random distribution, 
``Poisson 2'', constrained to fit the observed numbers of multi-planet systems (170) and planets within said systems (408).  The Poisson 2 random 
distribution fits the numbers of \ik systems with multiple transiting planets quite well, but it only accounts for $\sim 35\%$ of the single planet 
systems detected. As \ik does not detect all planets orbiting target stars for which some planets are observed, the observed multiplicities differ from the true multiplicity; we explore these differences in Section \ref{sec:coplanar}.

\begin{deluxetable*}{crrrrr}
\tabletypesize{\scriptsize}
\tablewidth{0pc}
\tablecaption{  \label{tab:Poisson} }
\tablehead{
\colhead{\# planets per system} & \colhead{Kepler} & \colhead{Poisson 0} & \colhead{Poisson 1} &\colhead{Poisson 2} & \colhead{RV (scaled to \ikt)} }
\startdata
\hline
0 		& 	159210	& 158976.5 			& 1630.9			& 227.3				& ---		\\
1 		&	791		& 1190.1				& 755.5			& 231.1				& 836.8	\\
2 		&	115		& 4.5					& 175.0			&117.6				& 97.2	\\
3 		&	45		& 1.1$\times 10^{-2}$	& 27.0			& 39.9				& 27.8 	\\
4 		&	8		& 2.1$\times 10^{-5}$	& 3.1				& 10.1				& 10.4	\\
5 		&	1		& 3.1$\times 10^{-8}$	& 2.9$\times 10^{-1}$& 2.1				& 3.5		\\
6 or more	&	1		& 3.9$\times 10^{-11}$	& 2.4$\times 10^{-2}$& 4.1$\times 10^{-1}$	& 3.5		\\
\hline
Total stars: 			&	\nall 	& \nall 	& 2591.9	& 628.4	& --- 		\\
Total planets:			&	\npl 	& \npl 	& \npl 	& 639.1	& \npl 	\\
Total stars with planets: 	&	\nsys	& 1194.5 	& \nsys 	& 401.1	& 981.0 
\enddata
\end{deluxetable*}

It is difficult to estimate the corresponding distribution for planets detected through Doppler radial velocity (RV) surveys, since we do not know how 
many stars were spectroscopically surveyed for planets for which none were found. However, a search through the Exoplanet Orbit 
Database \citep{Wright:2011} as of January 2011 shows that 17\% of the planetary systems detected by RV surveys are 
multi-planet systems (have at least two planets), compared to 21\% for \ikt. Including a linear RV trend as evidence for an additional planetary 
companion, \cite{Wright:2009} show that at least 28\% of known RV planetary systems are multiple. Note, however, that the radial velocity surveys are 
generally probing a population of more massive planets and one that includes longer period planets than the \ik sample discussed here. Some of the most 
recent RV planets are from the same population of small-planet, short-period multiples seen in the \ik data \citep[e.g.,][]{LoCurto:2010,Lovis:2011}. Also, the 
observed multiplicity of RV systems is far less sensitive to the mutual inclinations of planetary orbits than is the multiplicity of \ik candidates. 
The final column in Table~\ref{tab:Poisson} shows the distribution of radial velocity detected planets normalized to the total number of \ik 
candidates.

\begin{figure}
\includegraphics[width=0.5\textwidth]{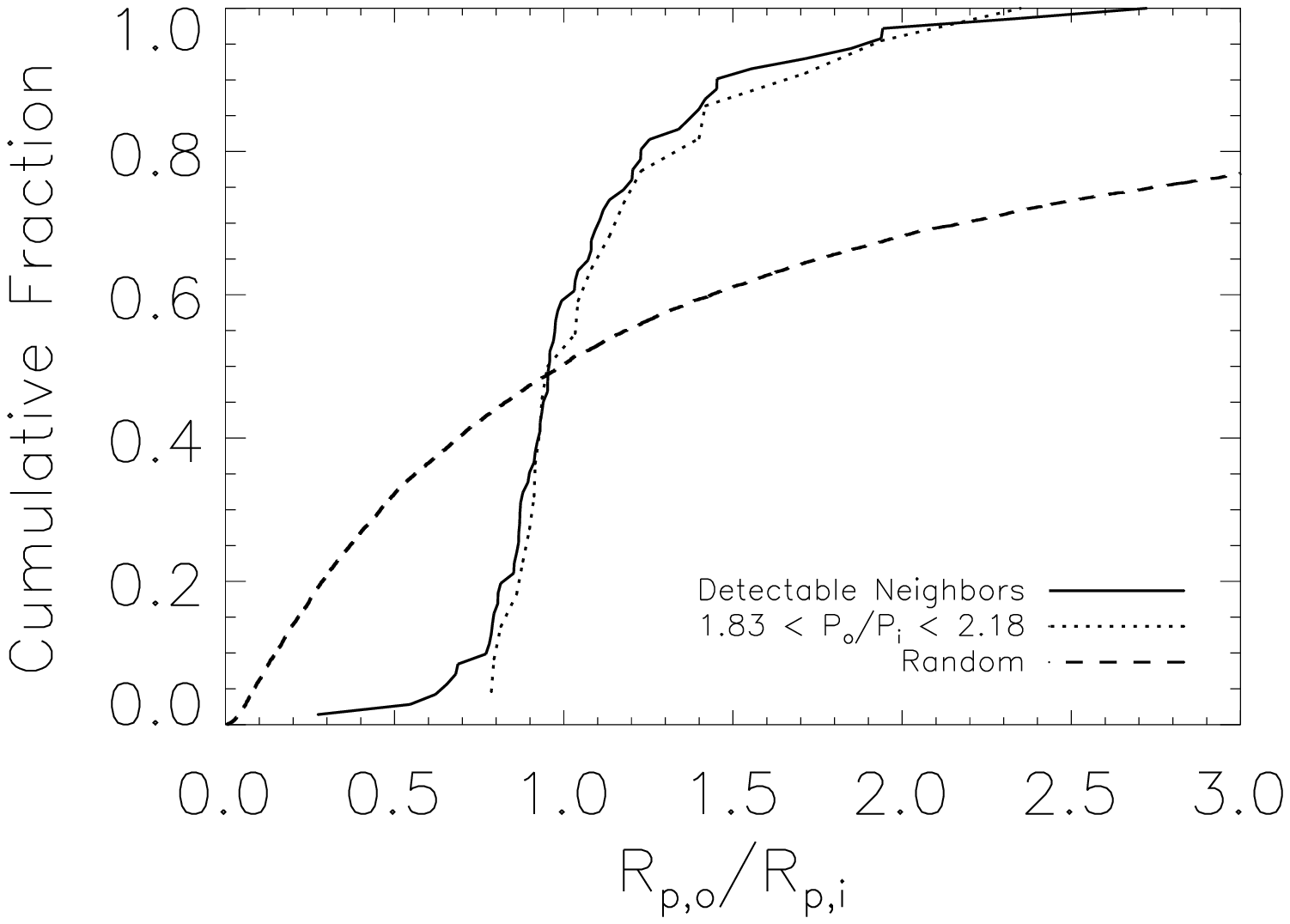}
\caption{Cumulative distribution of the ratio of planetary radii
($R_{\rm p,o}/R_{\rm p,i}$) for neighboring pairs of transiting
planet candidates for which both planets are detectable at the longer period.  The distribution of radii ratios for all
such neighboring pairs (solid line, 71 ratios) is the same (within statistical uncertainty) as that of such
pairs of planet candidates that are near the 2:1 mean motion resonance (dotted line, 22 ratios).  
Interestingly, neighboring planets tend to have very similar radii, with most of the population near $R_{\rm p,o}/R_{\rm p,i} \approx 1$. This strong 
tendency is 
illustrated by showing for contrast the cumulative distribution of ratios between radii randomly drawn from the debiased distribution (dashed line). In the debiased 
distribution, there is 
no significant preference for 
 $R_{\rm p,o}$ to be greater or less than $R_{\rm p,i}$. }
\label{RadiusRatioCum}
\end{figure}

In systems with multiple transiting candidates, the ratio(s) of planetary periods and the ratio(s) of planetary radii are well-measured independently of uncertainties in stellar 
properties. This suggests an investigation into whether outer planets are larger than inner planets on average, in order to provide constraints on 
theories of planet formation. (The ratio of orbital periods is discussed in Section \ref{sec:res}.) When considering the entire 
multi-candidate population, there is a slight but significant preference for outer planets to be larger (Figure~\ref{fig:perrad}). However, since planets with longer periods 
transit less frequently, all else being equal the SNR scales as $P^{-2/3}$, after accounting for the increase in SNR due to longer duration transits 
(assuming circular orbits). To debias the radius ratio distribution, each planetary system is investigated and if the smallest planet cannot be 
detected at SNR $> 16$ if it were placed in the longest period, that planet is removed. The remaining population of systems are not sensitive to 
observational bias, since each planet could be detected at all periods. The distribution of radius ratios ($R_{\rm p,o}/R_{\rm p,i}$), where the subscripts o and i refer to the inner and outer members of the pair of planets, for 
debiased 
planetary systems is shown by the solid curve in Figure \ref{RadiusRatioCum}. Interestingly, neighboring planets tend to have very similar radii, with most of the population near 
$R_{\rm p,o}/R_{\rm p,i} \approx 1$. This strong tendency is illustrated by comparing 
the cumulative distribution of observed radii ratios with ratios of  radii randomly 
drawn from the debiased distribution (dashed curve in Figure \ref{RadiusRatioCum}).  Since there is no requirement that the apparent depths of false positives and/or planets to be comparable, this suggests that false positives are not common among systems with radii ratios near unity. In the debiased distribution, there is no significant preference for
 the outer planet to be larger than the inner one.

Examination of this debiased sample for additional non-trivial trends and correlations (between radius ratio, period ratio, radius, period, 
multiplicity, near-resonance (defined by $\zeta_1$ discussed in Section 5), and stellar properties, reveals a few additional clues into the distribution 
of the \ik candidates. When there is a giant ($R > 6 R_{\oplus}$) planet, it is usually, but not always, at longer period. Radius ratios near 1 are 
almost exclusively found for planets less than 4 $R_{\oplus}$. Another trend is that whenever the period ratio between two neighboring detectable 
planets exceeds 5, the sizes between the planets were similar and, conversely, large radius ratios only occurred for period ratios $\lesssim$3. 
Furthermore, large period ratio systems are only found around small stars $(R_\star < R_{\odot})$. These latter examples may be due to observational 
bias.

\section{Reliability of the Sample}\label{sec:reli}

Very few of the planetary candidates presented herein have been validated or confirmed as true exoplanets.  As discussed in B11, \cite{Brown:2003} and \cite{Morton:2011}, the overwhelming 
majority of false positives are expected to come from eclipsing binary stars.  Many of these will be eclipses of fainter stars (physically associated to the target or background/faint foreground objects) within the 
aperture, others grazing eclipses of the target 
star. Similarly, transits of fainter stars that appear within the target 
aperture by large planets may also mimic small planets transiting the target star.  The vast 
majority of all active \ik planet candidates were identified using automated programs that do not discriminate between single 
candidates and systems of multiple planets. In addition, $\lesssim$ 1\% of single and multiple candidate systems were identified in 
subjective visual searches; this fraction is too small to significantly affect our statistical results.

The fact that planetary candidates are clustered around targets --- there are far more targets with more than one 
candidate than would be expected for randomly distributed candidates (Section \ref{sec:char}) --- suggests that the reliability of the 
multi-candidate systems is likely to be higher than that of the single candidates. We know from radial velocity observations 
that planets frequently come in multiple systems \citep{Wright:2009}, whereas astrophysical false positives are expected to be nearly random. 
The mere presence of multiple candidates increases our confidence that most or all are real planets, since the probability of 
multiple false positive signals is the product of the probabilities of two or more relatively rare cases \citep{Ragozzine:2010}. (Triple star systems would be dynamically unstable for 
most of the orbital period ratios observed.  Their dynamics would also give rise to very large eclipse timing variations 
\citep[e.g.,][]{Carter:2011}, which are generally not seen.)  Additionally, the concentration of candidate pairings with period ratio near first-order mean motion resonances such as 3:2 
and 2:1 (Figure~\ref{fig:cumprat} and \ref{fig:deriv}) suggests that these subsamples likely have an even larger fraction of true planets, since such 
concentrations would not be seen for random eclipsing binaries.  These qualitative factors have not yet been 
quantified.

Strong TTV signals are no more common for planets in observed multi-transiting candidates than for single candidates \citep{Ford:2011}, and it is known that some of these systems are actually stellar triples masquerading as planetary systems, suggesting that 
large TTVs are often the result of stellar systems masquerading as planetary systems \citep[e.g.,][]{Steffen:2011}.  Nonetheless, it is possible 
that weaker TTVs correlate with multi-transiting systems, but that these small TTVs  (e.g., those in 
the Kepler-11 planetary system, \citealt{Lissauer:2011}) require more data and very careful study to reveal.

The ratio of transit durations normalized to the observed periods of 
candidates in
multi-transiting systems can also be used to identify false
positives (generated by a blend of two objects eclipsing two stars of
different densities) or as a probe of the eccentricity and inclination
distributions.   Transit durations depend upon orbital period, and it is 
convenient to normalize values for candidates in a system by dividing 
measured duration by $P^{1/3}$; this normalization would yield the same 
value for planets of differing periods on circular orbits with the same 
impact parameter around the same star.
For this study, we assumed circular orbits and a random distribution of 
impact parameters as in \cite{Holman:2010a}. We find that the 
distribution of normalized
transit duration ratios is consistent with that of a population of 
planetary systems
with no contamination.  The consistency of normalized
duration ratios is not a
strong constraint for false positives, but the observed systems also pass 
this test.

There are five pairs for which the observed transit duration ratio
falls in the upper or lower 5\% of the distribution predicted.  These
are: KOI-864.01 and 864.03,  291.02 and 291.01, 1426.01 and 1426.03,
645.01 and 645.02 and 1089.02 and 1089.01; independent evidence from the 
shape of the transit light curve indicates that KOI-1426.03's transits 
are grazing (Section \ref{sec:stability}). However, this test is not definitive enough to rule out the planetary interpretation, in favor of a blended eclipsing binary hypothesis, for any of these pairs.
 Given
238 pairs of neighboring planets among the multiple planet candidate
systems identified by {\it Kepler}, a 10\% false alarm rate
(corresponding to the extreme 5\% on either side of the distribution)
could easily result in nearly two dozen such pairs of planets simply
due to chance.  The substantially smaller fraction of extreme normalized 
duration ratios that are observed may result from the lower SNR (and 
thus decreased likelihood for detection) of grazing and near-grazing 
transits.  In sum, we \emph{do not} find evidence for false
positives based on the ratio of transit durations of KOIs with a common
host star.

\section{Long-term Stability of Planetary Systems} \label{sec:stability}

{\it Kepler} measures planetary sizes, whereas masses are the key parameters for dynamical studies.  For our dynamical studies, we convert planetary radii, $R_p$, to masses, $M_p$, using 
the following simple formula: \begin{equation} M_p = \Big{(}\frac{R_p}{R_\oplus}\Big{)}^{2.06} M_\oplus ,\label{eq:mr} \end{equation} where R$_\oplus$ and M$_\oplus$ are the
 radius and mass of the Earth, respectively.  The power-law of Equation~(\ref{eq:mr}) was obtained by fitting to 
Earth and Saturn; it slightly overestimates the mass of Uranus (17.2 M$_\oplus$ vs.~14.5 M$_\oplus$) and slightly underestimates the 
mass of Neptune (16.2 M$_\oplus$ vs.~17.1 M$_\oplus$). (Note that Uranus is larger, but Neptune is more massive, so fitting both with a 
monotonic mass-radius relation is not possible.) Observations of transiting exoplanets \citep[][and references therein]{Lissauer:2011}  show more significant deviations from the relationship given by Equation (1) than do Uranus and Neptune, with both denser and less dense planets known, but on average the known exoplanets smaller than Saturn are consistent with this general trend.
The stellar mass, taken from B11, was 
derived from the $R_\star$ and log$~g$ estimates from color photometry of the \ik Input Catalog \citep{Brown:2011}.  

A convenient metric for the dynamical proximity of the $j^{th}$ and ($j+1$)$^{th}$ planets is the separation of their orbital semi-major axes, $a_{j+1}-a_j$, measured 
in units of their mutual Hill sphere radius  \citep{Hill:1878, Smith:2010}:
\begin{equation}
R_{H_{j,j+1}} = \Big{[}\frac{M_j + M_{j+1}}{3 M_\star} \Big{]}^{1/3} \frac{(a_j+a_{j+1})}{2}, \label{eq:rhill}
\end{equation}  
where the index $j = (1, N_p-1)$, with $N_p$ being the number of planets in the system under consideration, $M_j$ and $a_j$ are the planetary masses and semi-major axes, and $M_\star$ is the central star's mass (the masses of interior planets are likely to be much smaller than the uncertainty in the star's mass, and thus have been neglected in Equation \ref{eq:rhill}).  

The dynamics of two-planet systems are a special case of the 3-body problem that is amenable to analytic treatment and simple numerically-derived scaling formulas.  For instance, a pair of planets initially on coplanar circular orbits with dynamical orbital separation
\begin{equation}
\Delta \equiv \frac{a_{\rm o}-a_{\rm i}}{R_H} > 2 \sqrt{3} \approx 3.46  \label{eq:delta}
\end{equation}
can never develop crossing orbits, and are thus called ``Hill stable'' \citep{Gladman:1993}.  Dynamical orbital separations of all planet pairs in two-planet systems for stellar masses given in B11 and planetary masses given by Equation~(\ref{eq:mr}) are listed in Table~\ref{tab:two}.  We see that all two-planet systems obey the criterion given in Expression~(\ref{eq:delta}). This is additional evidence in favor of a low false positive rate, as random false positives would occasionally break this criterion.

The actual dynamical stability of planetary systems also depends on the eccentricities and mutual inclination \citep{Veras:2004}, none of which can be measured well from the transit data alone (cf.~\citealt{Ford:2011}).  Circular coplanar orbits have the lowest angular momentum deficit (AMD), and thus are the most stable \citep{Laplace:1784,Laskar:1997}. An exception exists for planet pairs protected by mean motion resonances (Subsection~\ref{sec:res191}) that produce libration of the relative longitudes of various combinations of orbital elements and prevent close approaches. A second exception exists for retrograde orbits \citep{Smith:2009}, but these are implausible on cosmogonic grounds \citep{Lissauer:2008}. Hence, the \ik data can only be used to suggest that a pair of planets is not stable. Instability is particularly likely if their semi-major axes are closer than the resonance overlap limit \citep{Wisdom:1980,Malhotra:1998}:
\begin{equation}
2\frac{a_{\rm o}-a_{\rm i}}{a_{\rm i}+a_{\rm o}} < 1.5 \Big{(}\frac{M_p}{M_\star}\Big{)}^{2/7}.
\end{equation}
Equation (4) was derived only for one massive planet; the other is massless.  Conversely, if it is assumed that both candidates are true planets, limits within the parameter space of planetary masses, eccentricities and mutual inclinations can be inferred from the requirement that the planetary systems are presumably stable for the age of the system, which is generally of order 10$^{11}$ times longer than the orbital timescale for the short-period planets that represent the bulk of \ikt's candidates.  However, we may turn the observational problem around: The ensemble of duration measurements can be used to address the eccentricity distribution \citep{Ford:2008}; since multi-planet systems have eccentricities constrained by stability requirements, they can be used as a check to those results \citep{Moorhead:2011}.  

Systems with more than two planets have additional dynamical complexity, but are very unlikely to be stable unless each neighboring pair of planets satisfies Expression~(\ref{eq:delta}).  Nominal dynamical separations ($\Delta$'s) between neighboring planet pairs in three-planet systems are listed in Table~\ref{tab:three}, the separations in four-planet systems are listed in Table~\ref{tab:four}, and those for the five-planet and six-planet systems in Table~\ref{tab:fiveplus}.  We find that in the overwhelming majority of these cases, the inequality in Expression~(\ref{eq:delta}) is satisfied by quite a wide factor.  

\cite{Smith:2009} conducted suites of numerical integrations to examine the stability of systems consisting of equal mass planets that were equally spaced in terms of $\Delta$.  They demonstrated dynamical survival of systems with three comparably-spaced 1 M$_\oplus$ planets orbiting a 1 M$_\odot$ star for $10^{10}$ orbits of the inner planet in cases where the relative spacing between orbital semi-major axes exceeded a critical number ($\Delta_{\rm crit} \sim 7$) of mutual Hill spheres. For systems with five comparably-spaced 1 M$_\oplus$ planets, they found $\Delta_{\rm crit} \approx 9$.  Their calculations give slightly larger $\Delta_{\rm crit}$ for systems of five 0.33 M$_\oplus$ planets, suggesting a slightly smaller $\Delta_{\rm crit}$ for the somewhat larger planets that dominate the \ik sample of interest here.  Systems with planets of substantially differing masses tend to be less stable because a smaller fraction of the AMD needs to be placed in the lowest mass planets for orbits to cross.  Taking all of these factors, as well as the likely presence of undetected planets in many of the systems, into account, we consider $\Delta_{\rm crit} \approx 9$ to be a rough estimate of how close planetary orbits can be to have a reasonable likelihood of survival on Gyr time scales.

\begin{figure}
\includegraphics[width=0.5\textwidth]{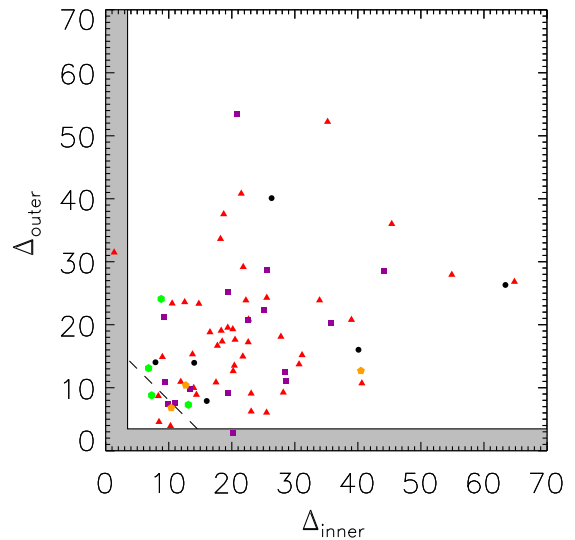}
\caption{ The orbital separations, expressed in mutual Hill radii (Equation~\ref{eq:rhill}), for the inner ($\Delta_{\rm i}$) and outer ($\Delta_{\rm o}$) pairs within 3-planet systems and adjacent 3-planet sub-systems of the 4-, 5-, and 6-planet systems of our sample.  The colored symbols are the same as in Figure~\ref{fig:perrad}, and we also include the spacings of planets within the Solar System as black circles.  Gray region: stability boundary for 2-planet systems, $\Delta>2\sqrt{3}$ (Expression~\ref{eq:delta}).  Dashed line: stability boundary used for the simulated population of multi-planet systems: $\Delta_{\rm i}+\Delta_{\rm o}>18$; some of the nominal systems survived long-term integrations despite transgressing that boundary.  The unstable nominal systems KOI-191 and KOI-248 both lie within the gray region, and all the long-term survivors are outside of it. }
\label{fig:hill}
\end{figure}

  For each adjacent set of 3 planets in our sample, we plot the $\Delta$ separations of both the inner and the outer pairs in Figure~\ref{fig:hill}.  We find that in some cases $\Delta < 9$, but the other separation is quite a bit larger than $\Delta_{\rm crit}$.  In our simulations of a population of systems in Section~\ref{sec:coplanar}, we impose the stability boundary: 
\begin{equation}  
\Delta_{j+1, j} + \Delta_{j+2, j+1} > 18,  \label{eq:2delta}
\end{equation}
which accounts for this possibility, in addition to requiring that inequality in Expression (\ref{eq:delta}) is satisfied. The form of Expression (\ref{eq:2delta}) is motivated by observations within our Solar System \citep{Lissauer:1995}.

We investigated long-term stability of all 55 systems with three or more planets using the hybrid integrator within the 
\emph{Mercury} package \citep{Chambers:1999}.  We set the switchover at 3 Hill radii, but in practice we aborted simulations that violated this limit, so for 
the bulk of the simulation the $n$-body mapping of \cite{Wisdom:1991} was used, with a time step of 0.05 
times as large as the orbital period of the innermost planet.  The simplest implementation \citep{Nobili:1986} of 
general relativistic precession was used, an additional potential 
\begin{equation}
U_{\rm{GR}} = -3\Big{(}\frac{GM_\star}{cr}\Big{)}^2, 
\end{equation}
where $G$ is 
Newton's constant, $c$ is the speed of light, and $r$ is the instantaneous distance from the star.  More sophisticated 
treatments of general relativity \citep{Saha:1994} are not yet required, due to the uncertainties of the masses of the planets and stars whose 
dynamics are being modeled.  We neglected precession due to tides on or rotational flattening in the planets, which are only significant 
compared to general relativity for Jupiter-size planets in few-day orbits \citep{Ragozzine:2009}, and precession due to the rotational oblateness of 
the star, which can be significant for very close-in planets of any mass.  Precession due to the time-variable flattening of the star can generate a 
secular resonance, compromising stability \citep{Nagasawa:2005}; we neglected this effect, treating the star as a point mass.  We also neglect 
tidal damping of eccentricities, which can act to stabilize systems over long time scales, and tidal evolution of semi-major axes, which sometimes has 
a destabilizing effect.  We assumed initially circular and coplanar orbits that matched the observed periods and phases, and chose the nominal masses of Equation~(\ref{eq:mr}).

For the triples, quadruples and the five-planet system KOI-500, we ran these integrations for $10^{10}$ orbital periods of the innermost planet, and found the nominal system to be 
stable for this span in nearly all cases; the two exceptions are described below.  The most populous transiting planetary system discovered 
so far is the six-planet system Kepler-11.  In the discovery paper, \cite{Lissauer:2011} reported a circular model, with transit time variations 
yielding an estimate of the masses of the five inner planets.  They also proposed two more models with planets on moderate eccentricities, and slightly 
different masses, which also fit the transit timing signals.    Continued integration showed that both of those eccentric solutions are nominally unstable, at 610~Myr for the 
all-eccentric 
fit, and at 427~Myr for an integration that was a very slight offset from the b/c-eccentric fit.  The all-circular fit remained stable for the duration of our 1~Gyr integration, so we conclude that the fits of that paper are physically plausible. Introducing a small amount of eccentricity can compromise stability, but our results do not exclude eccentricities of a few percent because (1) planetary masses have significant uncertainties, and we have not conducted a survey of allowed mass/eccentricity combinations, and (2) tidal damping of eccentricities may be a significant stabilizing mechanism over these time scales for planets orbiting  as close to their star as are Kepler-11b and c.  Future work should address the limits imposed on the eccentricities by the requirement that the system remain stable for several Gyr.  

Only two systems became unstable at the nominal masses: KOI-284 and KOI-191; KOI-191 is discussed in Subsection~\ref{sec:res191}. KOI-284 has a pair of candidates with periods near 6 days and a period ratio $1.0383$ and an additional planet at 18 days.  Both of the 6-day planets would need to have 
masses about that of the Earth or 
smaller for the system to satisfy Expression~(\ref{eq:delta}) and thus be Hill stable.  But the vetting flags are 3 for both of these 
candidates --- meaning that they are suspect candidates for other reasons --- and the \ik data display significant correlated noise that may be 
responsible for these detections.  We expect one or both of these candidates are not planets, or if they are, they are not orbiting the 
same star. This is the only clear example of a dynamically-identified likely false positive in the entire \ik multi-candidate population.

In our investigation, we also were able to correct a poorly fit radius of 35~R$_{\oplus}$ (accompanied by a grazing impact parameter 
in the original model, trying to account for a ``V'' shape) for KOI-1426.03, which resulted in too large a nominal mass, driving the 
system unstable.  It is remarkable that stability considerations allowed us to identify a poorly-conditioned light curve fit.  Once the 
correction was made to 13~R$_{\oplus}$, the nominal mass allowed the system to be stable.

That so few of the systems failed basic stability measurements implies that there is not a requirement that a large number of planets have 
substantially smaller densities than given by the simple formula of 
Equation (\ref{eq:mr}). Stability constraints have the potential  
to place upper limits on the masses and densities of candidates in multiply transiting systems, assuming that all the candidates are planets.

\section{Resonances} \label{sec:res}

The formation and evolution of planetary systems can lead to preferred occupation of resonant and near-resonant configurations \citep{Goldreich:1965, Peale:1976, Malhotra:1998}.  The abundance of resonant planetary systems provides constraints on models of planetary formation and on the magnitude of differential orbital migration.

The distribution of period ratios of multi-transiting candidates in the same system is significantly different from the distribution obtained by taking ratios of 
randomly selected pairs of periods from all 408 multi-transiting candidates. Most observed planetary pairs are neither in nor very near low-order mean motion resonances; nonetheless, the number of planetary pairs in or near mean motion resonances exceeds that of a random distribution (Figures  \ref{fig:cumprat} -- \ref{fig:deriv}, \ref{fig:zetastatistic} -- \ref{fig:kstest}). In this section, we quantify this clustering and discuss a few particularly interesting candidate resonant systems.

\subsection{Resonance Abundance Analysis}

\ikt's multi-planet systems provide a complementary data set to exoplanets detected in RV surveys, in that RV planets are typically larger and detected 
on the basis of mass rather than radius. Radial velocity multiples have the advantage of being minimally biased by relative planetary inclinations. 
Multiply-transiting planets have two advantages over RV planetary systems for the study of resonances: First, periods are measured to very high 
accuracy, even for systems with moderately low SNR, so period ratios can be confidently determined without the harmonic ambiguities that affect RV 
detections \citep{Anglada-Escude:2010, Dawson:2010}; however, transits can be missed or noise can be mistaken for additional transits when the SNR is 
very low. Second, resonances significantly enhance the signal of transit timing variations \citep{Agol:2005, Holman:2005, Veras:2010}, which can also 
be measured accurately, either to demonstrate or confidently exclude resonance occupation, given enough time \citep[e.g.,][]{Holman:2010a}.

Planets can be librating in a resonance even when they have apparent periods that are not perfectly commensurate 
\citep{Peale:1976, Marcy:2001, Rivera:2001, Ragozzine:2010}. Nevertheless, the period ratios of resonant planets will generally be 
within a few percent of commensurability. In the cumulative 
distribution of period ratios (Figure \ref{fig:cumprat}), we see that few planets appear to be directly in the resonance (with the 
important exception of KOI-730), but placing limits on resonance occupation requires additional analysis.

\begin{figure}
\begin{center}
\includegraphics[width=0.5\textwidth]{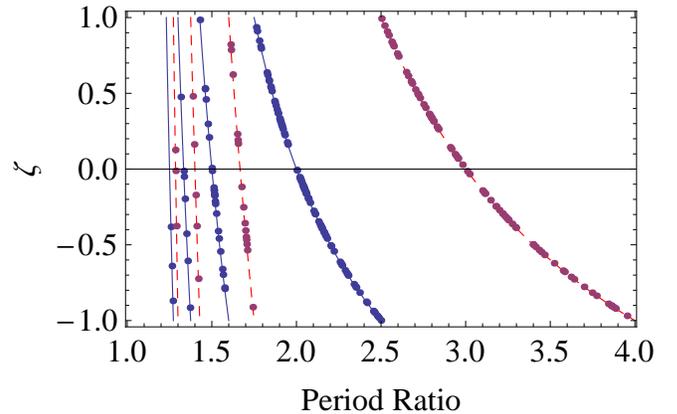}
\caption{The value of $\zeta$ as a function of period ratio for the first order (solid, blue) and second order (dashed, red) MMRs.  
Also shown are the observed period ratios in the \ik data.\label{fig:zetastatistic}}
\end{center}
\end{figure}

In the Solar System, there is a known excess of satellite pairs near mean-motion resonance (MMR) \citep{Goldreich:1965}, and theoretical 
models of planet formation and migration suggest that this may be the case for exoplanets (e.g., \citealt{Marcy:2001,Terquem:2007}).  Can such 
an excess be seen in the sample of multiply transiting \ik planetary systems?  To answer this question, we divide the period ratios of the 
multiply transiting systems into ``neighborhoods''  that surround the first and second-order ($j$:$j-1$ and $j$:$j-2$) MMRs.  The boundaries between 
these neighborhoods are chosen at the intermediate, third-order MMRs.  Thus, all candidate systems with period ratios less than 4:1 
are in some neighborhood.  For example, the neighborhood of the 3:1 MMR runs between period ratios of 5:2 and 4:1, for the 2:1 it runs 
from 7:4 to 5:2.  With this algorithm,  the first-order MMRs have neighborhoods (represented by solid lines in Figure \ref{fig:zetastatistic}) that are essentially twice as large as those for 
second-order MMRs (dashed lines in Figure \ref{fig:zetastatistic}). The purpose of dividing the period ratio distribution into neighborhoods around the major resonances is to account for the fact that even random period ratios can be close to some ratio of integers \citep{Goldreich:1965}.

We define a variable, $\zeta$, that is a measure of the difference between an observed period ratio and the MMR in its neighborhood.  
In order to treat all neighborhoods equally,  $\zeta$ is scaled to run from $-1$ to $1$ in each neighborhood.  For first-order MMRs, $\zeta$ is given by
\begin{equation}
\zeta_1 \equiv 3 \left( \frac{1}{{\cal P} - 1} - \rm{Round}\left(\frac{1}{{\cal P} - 1}\right) \right) ,
\end{equation}
where ${\cal P} \equiv P_{\rm o}/P_{\rm i}$ is the observed period ratio (always greater than unity) and ``Round'' is the standard rounding function that returns the 
integer nearest to its argument.  Similarly, for second-order MMRs, $\zeta$ is given by
\begin{equation}
\zeta_2 \equiv 3 \left( \frac{2}{{\cal P} - 1} - \rm{Round}\left(\frac{2}{{\cal P} - 1}\right) \right).
\end{equation}

Table \ref{tab:bincounts} gives the number of planet pairs taken from Tables~\ref{tab:two}--\ref{tab:fiveplus} that are found in each neighborhood.  The value of $\zeta$ as a function of the period ratio is shown together with the data in Figure \ref{fig:zetastatistic}. We also list neighborhood abundances
considering only adjacent planet pairs.  This last case primarily removes pairs in the neighborhood of the 3:1 MMR, but also a nontrivial fraction of those neighboring the 2:1 MMR.  The fact that significant numbers of non-adjacent pairs are in these neighborhoods demonstrates how closely packed many of the \ik multi-transiting systems are.

\begin{table}
\begin{center}
\caption{Counts of planet pairs found in each resonance neighborhood.\label{tab:bincounts}}
\begin{tabular}{ccc} \hline \hline
MMR & \# Total Pairs & \# Adjacent Pairs\\ \hline
2:1 & 89 & 78 \\
3:2 & 22 & 22 \\
4:3 & 7  & 7  \\
5:4 & 3  & 3  \\ \hline
3:1 & 79 & 55 \\
5:3 & 15 & 15 \\
7:5 & 5  & 5  \\
9:7 & 3  & 3  \\ \hline
\end{tabular}
\end{center}
\end{table}

\begin{figure}
\begin{center}
\includegraphics[width=0.4\textwidth]{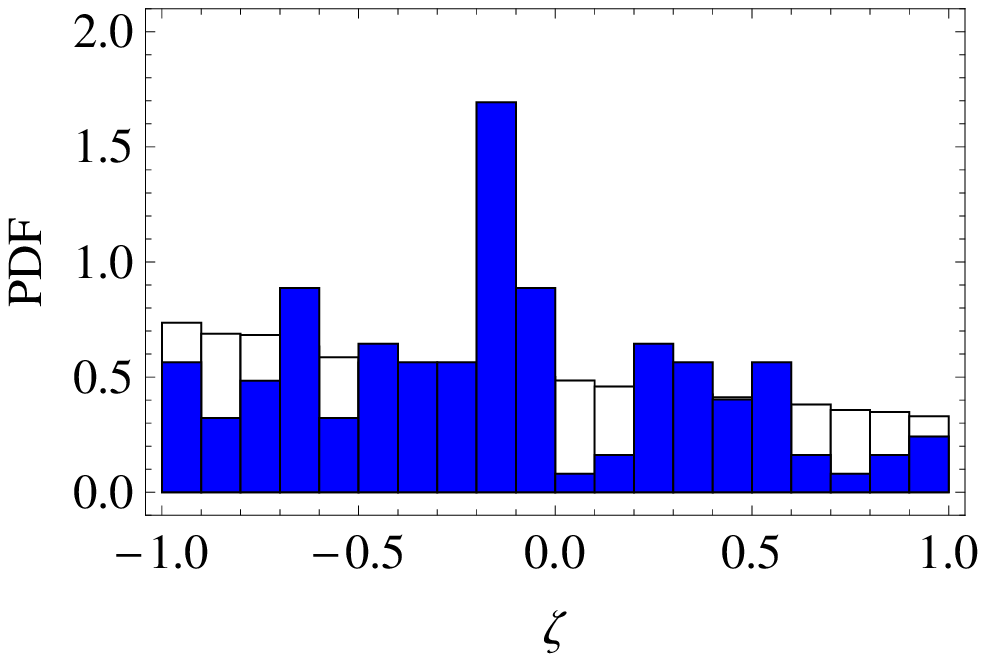}\\
\includegraphics[width=0.4\textwidth]{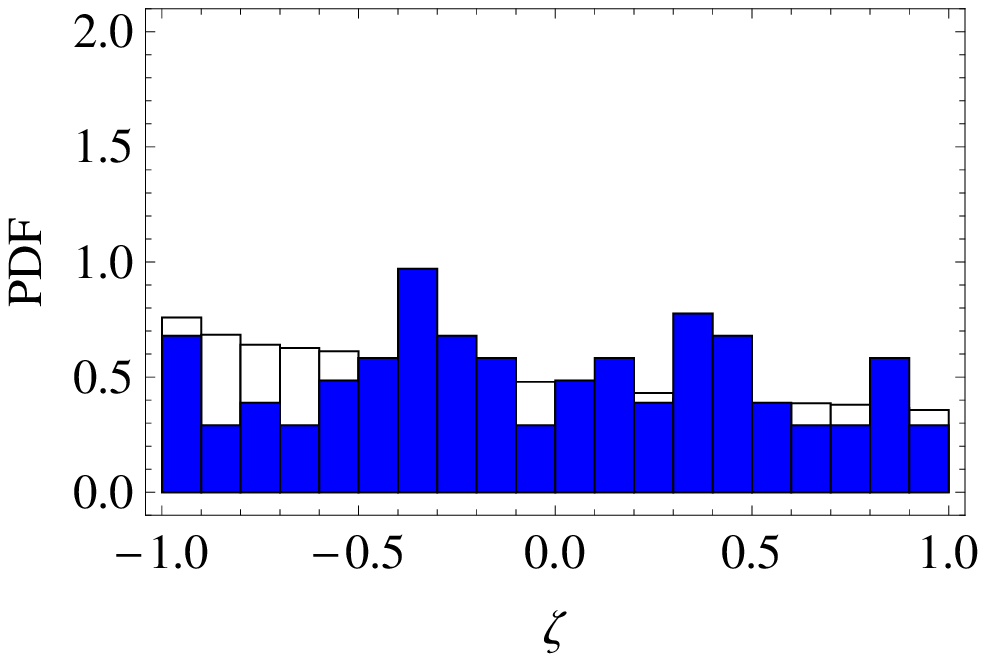}
\caption{Probability density of the systems as a function of $\zeta$ for first order (top) and second order (bottom) mean motion resonances.  The most common 
values for $\zeta$ for first-order resonances are small and negative, i.e., lie just outside the corresponding MMR, while one of the least common values, small and positive, lie just inside the 
MMR.  No strong trends are observed for second-order resonances.  The probability density for the logarithmically-distributed test sample is  a monotonically decreasing function and is shown for reference in both plots by white boxes for those values of $\zeta$ for which its value exceeds that of the data.\label{pdfs}}
\end{center}
\end{figure}

We calculate $\zeta$ for each planetary period ratio and stack the results for all of the resonances in each MMR order.  The resulting distribution in $\zeta$ is 
compared to a randomly-generated sample drawn from a uniform distribution in $\log {\cal P}$, a uniform distribution in ${\cal P}$, and from a 
sample constructed by taking the ratios of a random set of the observed periods in the multiple candidate systems.  This third sample 
has a distribution that is very similar to the logarithmic distribution.  Figure \ref{pdfs} shows a histogram of the number of systems 
as a function of $\zeta$ for both the first-order and second-order MMRs.  We see that the results for the second-order MMRs have some qualitative 
similarities to the first-order MMR, though the peaks and valleys are less prominent.  In the tests described below, we also consider the case where all first and second-order MMRs are stacked together, as well as this stacked combination with the likely resonant systems KOIs 730, 191 and 500 (Subsections \ref{sec:res730}, \ref{sec:res191}, and \ref{sec:res500}) removed in order to determine the robustness of our results to the influence of a few special systems.  Finally, we study the distribution of adjacent planets only. 

\begin{figure}
\begin{center}
\includegraphics[width=0.4\textwidth]{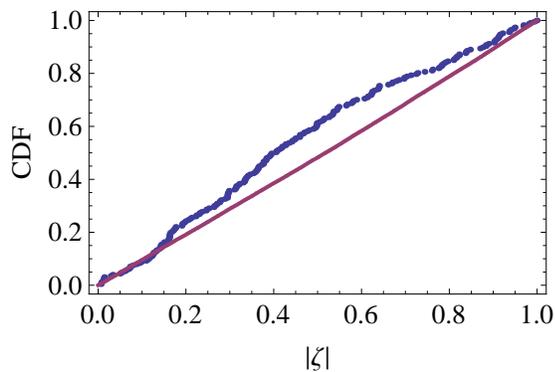}
\caption{Cumulative distribution function (CDF) of the absolute value of $\zeta$ for the combined first-order and second-order resonances (blue).  Also shown is the CDF for the logarithmically distributed test sample (red).  The results of the KS test for these two distributions are given in Table \ref{tab:ksresults} along with the KS test results for other combinations of resonances and test distributions.
\label{fig:kstest}}
\end{center}
\end{figure}

To test whether the observed distributions in $\zeta$ are consistent with those from our various test samples, we took the absolute 
value of $\zeta$ for each collection of neighborhoods and used the Kolmogorov-Smirnov test (KS test) to determine the probability that 
the observed sample is drawn from the same distribution as our test sample.  The results of these tests are shown in Table 
\ref{tab:ksresults}.  Figure \ref{fig:kstest} shows that  the probability density function (PDF) for the combined distribution of first and second-order MMRs differs from the PDF for the logarithmically-distributed sample.

\begin{table*}
\begin{center}
\caption{KS test p-values for different resonance sets and for test distributions.\label{tab:ksresults}}
\begin{tabular}{clll} \hline \hline
Resonance set               &$\log(\cal P)$& $\cal P$     & All \ik multis \\ \hline
2:1 only                    & 0.00099  & 0.00011  & 0.00059  \\
$j$:$j-1$                   & 0.0012 & 0.00021 & 0.00044 \\
$j$:$j-2$                   & 0.046   & 0.0094   & 0.040   \\
all ($j$:$j-1$ and $j$:$j-2$) & $8.2\times 10^{-5}$ &$2.7\times 10^{-7}$&$3.9\times 10^{-5}$  \\
all except KOIs 191, 500, 730 & 0.0010  & 0.00014 & 0.00072 \\
all adjacent pairs & $9.7\times 10^{-6}$ & $7.5\times 10^{-7}$ & $7.5\times 10^{-6}$ \\ \hline
\end{tabular}
\end{center}
\end{table*}

A few notable results of these tests include: (1) the distributions observed within the combined first-order and second-order neighborhoods are distinctly inconsistent with any 
of the trial period distributions---including a restricted analysis where the likely resonant systems KOIs 191, 500, and 730 are excluded, (2) the distributions within the neighborhoods surrounding the first-order MMRs are also very unlikely to originate from the test distributions, (3) there is significant evidence that the distributions within the neighborhood surrounding the 2:1 MMR alone is inconsistent with the test 
distributions, (4) there is a hint that the distributions in the neighborhoods of second-order resonances are not consistent with the test distributions, particularly the test
distribution that is uniform in $\cal P$.  Additional data are necessary to produce higher significance results for the second-order 
resonances, and (5) when only adjacent planet pairs are considered, all of the test distributions are rejected with higher significance.

Given the distinct differences between our quasi-random test distributions and the observed distribution, a careful look at the histograms shown in 
Figure \ref{pdfs} reveals that the most common location for a pair of planets to reside is slightly exterior to the MMR (the planets are 
farther apart than the resonance location), regardless of whether the resonance is first-order or second-order.  Also, a slight majority (just 
over 60\%) of the period ratios have $\zeta$ values between --1/2 and +1/2, while all of the test distributions have roughly 50\% 
between these two values in $\zeta$ (note that $\zeta = \pm 1/2$ corresponds to sixth-order resonances in first-order neighborhoods and 
twelfth-order resonances in second-order neighborhoods---none of which are likely to be strong for these planet pairs).  There are very few examples of systems with planet pairs that lie slightly closer to each other than the first-order 
resonances, and for the second-order resonances only the 3:1 MMR has planet pairs just interior to it.  Finally, there is a hint that 
the planets near the 2:1 MMR have a wider range of orbital periods, while those near the 3:2 appear to have shorter periods and smaller 
radii on average. While additional data are necessary 
to claim these statements with high confidence, the observations are nonetheless interesting and merit serious investigation to determine 
the mechanisms that might produce such distributions.

A preference for resonant period ratios cannot be exhibited by systems without direct dynamical interaction. Multiple star systems with similar orbital periods (i.e., non-hierarchical) are not dynamically stable, so dynamical interaction between objects with period ratios 
$\lesssim$ 5 indicates low masses, typically in the planetary or brown dwarf regime. Thus, the preference for 
periods near resonances in our dataset is statistically significant evidence that most if not all of the candidate (near-)resonant systems are actually systems of two or 
more planets.  Note, however, that assessing the probability of dynamical proximity to resonance for the purpose of 
eliminating the false positive hypothesis in any particular system requires a specific investigation.

An excess of pairs of planets with separations slightly wider than nominal resonances was predicted by \citet{Terquem:2007}, who pointed out that 
tidal 
interactions with the star will, under some circumstances, break resonances. However, not all of the systems near resonance have short-period planets 
that would be strongly affected by tides. 
Interestingly, KOI-730 seems to have maintained stability and resonance occupation despite the short periods of the planets that imply susceptibility of the system to differential tidal evolution.

\subsection{Frequency of Resonant Systems} \label{secFreqRes}

Radial velocity planet searches suggest that roughly one-third of the multiple planet systems that have been well-characterized by RV observations are 
near a low-order period commensurability, with one-sixth near the 2:1 MMR \citep{Wright:2011}.  \ik observations find that at least $\sim~16\%$ of 
multiple transiting planet candidate systems contain at least one pair of transiting planets close to a 2:1 period commensurability ($1.83 < ~{\cal P} 
< 2.18$).  While it is tempting to consider these results as showing a similarity between the Jupiter-mass planets that dominate the RV sample and 
those of the 
Neptune-size planets that form the bulk of the \ik candidates, there are many differences in the criteria used to identify these planets and to 
calculate the period ratios. 

Radial velocity planets are spread over a much wider range of period ratio than are the \ik candidates, and the concentration seen in the RV planets is in the 
region very near the 2:1 MMR, whereas the one-sixth number for the \ik candidates is in the wider ``neighborhood'' that we have defined. 
\cite{Wright:2011}'s estimate for RV systems near the 2:1 MMR may be lower than the actual value for the ensemble of systems that they considered 
because of detection biases.  
Radial velocity observations for systems near the 2:1 MMR have an approximate degeneracy with a single planet on an eccentric orbit.  This degeneracy 
makes it difficult for RV observations to detect a low-mass planet in the interior location of a 2:1 MMR \citep{Giuppone:2009,Anglada-Escude:2010}.

  Among RV-discovered systems, only one-quarter of the pairs near
the 2:1 MMR have an inner planet significantly less massive than the
outer planet.  The two exceptions (GJ 876 c\&b, $\mu$ Ara d\&b) each
benefited from a unusually large ($\gtrsim 100$) numbers of RV
observations.  Because of this degeneracy, it is possible that many 2:1 resonant systems with low-mass, 
interior members have been misidentified as 
eccentric single planets.  Thus, the true rate of planetary
systems near the 2:1 MMR may be significantly greater than one-sixth.

Also, the true fraction of \ik systems
near a 2:1 period commensurability could be significantly greater than
$\sim~16\%$, since not all planets will transit.  If we assume that
planets near the 2:1 MMR are in the low inclination regime, then the
outer planet should transit about ${a_1}/{a_2} \approx 63\%$ of the time, implying that
the true rate of detectable planets (in size-period space, not accounting for the geometrical limitations of transit photometry) near the 2:1 MMR is  $\gtrsim~25\%$.  If a significant
fraction of these systems are not in the low inclination regime, then
the true rate of pairs of planets near the 2:1 MMR would be even
larger.

Based on the B11 catalog, neighboring transiting planet candidates tend to have similar radii, and in the majority of cases, the outer planet is 
slightly larger.  However, smaller planets are more easily detected in shorter-period orbits, and a debiased distribution shows no preference for the outer planet to be larger (Section \ref{sec:char}).
The distribution of planetary radii ratios ($R_{\rm p,o}/R_{\rm
p,i}$) for neighboring pairs of transiting planet candidates near a
2:1 period commensurability (mean motion resonance, MMR) is concentrated between 0.8 and 1.25,  (see Figure \ref{RadiusRatioCum},
dotted curve).  For reasonable assumptions of a mass-radius relationship (Equation
\ref{eq:mr}), this implies that most neighboring transiting
planet candidate systems have masses within $\sim$40\% of each other.
Fortunately, radial velocity surveys can distinguish between pairs of
planets with similar masses and a single eccentric planet.  However,
those pairs in the tail
($\sim~20\%$) for which the radii ratio exceeds 1.4
could be difficult to detect with radial velocity observations, since for nominal mass-radius values,
the RV signatures of these pairs would differ from that of a
single eccentric planet by less than 30\% of the velocity amplitude of
the outer planet (on the decadal timescales that are typical of RV observations).
 Based on the planet radii ratio distribution from \ikt, we estimate
that the abundance of neighboring planets near the 2:1 MMR could be
$\sim~20$\% greater than suggested by radial velocity surveys due to the
difficulty in distinguishing the RV signature of such systems
from a single eccentric planet. A more precise determination of the frequency of 
resonant or near-resonant systems based on \ik and/or RV surveys is left for future work.

\subsection{KOI-730: A Multiply-Resonant Candidate System} 
\label{sec:res730}

While few nearly exact mean motion resonances are evident in the sample 
of \ik planetary candidates, one system stands
out as exceptional: The periods of the four candidates in KOI-730 
satisfy the ratio 8:6:4:3 to $\sim$ 1 part in 1000 or
better. This resonant chain is potentially the missing link that explains how 
planets that are subject to migration in a gas or planetesimal disk can avoid 
close encounters with each other, being brought to a very closely-packed, yet 
stable, configuration.  Mechanisms that gently ease planetary pairs out of such 
resonant configurations probably account for the observed preference for pairs to 
be just wide of resonances, as found above.  

This system is difficult to study because of the faintness of the target star (\ik magnitude $Kp$ = 15.34).  This faintness, combined with the small 
transit depths means individual transits are only marginally detected. Indeed, this star was not even observed during Q4 because it was considered a 
marginal target until the first of its planet candidates was identified.  Also, its location within the sky is such that during autumn quarters (Q3, 
Q7, Q11,...) its light falls upon \ikt's CCD Module 3, which failed early in Q4, so it can only be observed by \ik about 70\% of the time\footnote{The 
\ik spacecraft rotates four times per orbit to keep the sunshade and solar panels oriented properly.  Targets are imaged on different parts of the 
focal plane during different orientations.  The \ik orbital period is $\sim$372 days, and the data are grouped according to the ``quarter'' year during 
which observations were made.  Regular observations were commenced about 65\% of the way through an orientation; the standard data taken prior to 
\ikt's first ``roll" are referred to as Q1, and the 10 days of early observations done for bright targets are called Q0.  Subsequent quarters are 
numbered sequentially: Q2, Q3, ...}.  The orbital period of planet candidate 730.03 was initially (B11) thought be half its presently-reported value, 
putting it in a 1:1 resonance with 730.02.  Our current best solution discounts every other transit of that solution, placing these two planets 
within the 2:1 resonance.

 KOI-730.03 shows some sign 
of weak TTVs through Q2 \citep{Ford:2011}. The remarkably commensurate 
period ratios of these four candidates give us strong 
confidence that they all will eventually be confirmed as planets orbiting the same star. We are working on a thorough analysis of 
this system, including transit-timing fits and long-term 
stability.

\subsection{KOI-191: A System Protected by Resonance?} \label{sec:res191}

KOI-191 is one of the first five \ikt\ targets announced as  showing multiple transiting exoplanet 
candidates \citep{Borucki:2010a, Steffen:2010}.  Since that time two additional planets have been seen.  Two planets in the KOI-191 system have a 
period ratio of $1.258$, and the estimated Hill separation between these two planets is $\Delta = 2.9$ (Table \ref{tab:four}).  One of these planets has a radius $R_p \sim 1$ R$_{\rm Jup}$.  Its mass would need to be $\lesssim 17$ 
M$_{\oplus}$ to satisfy Expression~(\ref{eq:delta}).  Alternatively, the two planets could be locked in 5:4 mean-motion resonance 
that prevents close approaches even if orbits cross.  Neptune and Pluto are locked in 3:2 mean-motion resonance of this type 
\citep{Cohen:1965}.  Relatively fast precession, on the order of 100 orbital periods, would be needed to maintain the resonance libration, as the 
observed 
period ratio differs from the nominal resonance location by 0.6\%.  We note that the precession time scale of the resonant pair of 
giant planets orbiting GJ 876 is tens of orbital periods \citep{Marcy:2001}. In short, the dynamical stability of this system 
merits further investigation.

\subsection{KOI-500: (Near-)Resonant Five-Candidate System} \label{sec:res500}   

KOI-500 is a 5-candidate system with periods 0.986779 d, 3.072166 d, 4.645353 d, 7.053478 d, and 9.521696 d. Neighboring pairs of 
the outer four of these planets all have period ratios about 1$\%$ greater than those of first-order two-body MMRs. Thus they are 
nearly commensurate, but unless there is an unexpectedly large amount (given the small planetary sizes, see 
Table \ref{tab:fiveplus}) of apse precession, they are not locked in librating two-body resonances.  Perhaps this is a consequence of diverging tidal 
evolution, as predicted by  Papaloizou \& Terquem (2010).   The three  adjacent period spacings between these four planets 
contribute to the statistics of period ratios just wide of resonance (small, negative $\zeta$).  

Of particular interest, the combinations of mean motions $2n_2 - 5n_3 +3n_4 \approx 1.6 \times 10^{-5}$ and $2n_3 - 6n_4 +4n_5 
\approx 1.3 \times 10^{-5}$ are so small (smaller than the uncertainties in measured values, i.e., consistent with 0) that we 
suspect two 3-body Laplace-like resonances and/or a 4-planet resonance may be active and controlling the dynamics of this system.  

\subsection{KOI-738 and KOI-787: Planets with Period Ratios of 9:7}\label{sec:res738}

The pair of planets in KOI-738 have an observed period ratio ${\cal P} = 1.285871$, and the pair in KOI-787 have  ${\cal P} = 
1.284008$  (Table \ref{tab:two}).  The ratio 9/7 = 1.285714..., so KOI-738 is less than 1 part in 5000 from exact  9:7 resonance 
and KOI-787 is less than 1 part in 500 from this resonance.  The 9:7 resonance is fairly strong if one or both planets have a 
nontrivial $e$ or if their orbits are inclined to one another.

\subsection{KOI-657 and KOI-812: Systems where Resonances Suggest Missing Planets}\label{sec:res657}

Many \ik planets likely have resonant companions that have not been detected because they do not transit or are too small.  Such 
unseen planets may ultimately be detected via TTVs that they induce on observed transiting planets \citep{Ford:2011}. 
Additionally, if they are members of resonant chains similar to that seen in KOI-730, the period ratio of two observed transiting planets in the chain 
may indicate the presence of missing links. If enough chains are ultimately identified, the fraction with missing links would yield constraints on the degree of non-planarity of multiply-resonant planetary systems. We investigate evidence for such chains in this subsection.

The strongest case for a missing link can be made if only one unseen planet is needed to complete the chain and the links provided 
by that planet would be much stronger than those without it.  Neighboring planets in KOI-730 have period ratios close to 4:3 or 
3:2; the resonances which correspond to the most significant spikes in the distribution of period ratios are 2:1 and 3:2; 
combinations of these period ratios can produce two-link chains with period ratios of the non-neighboring planets of 16/9, 2, 9/4, 
8/3, 3 or 4.  As 2:1 is a first-order resonance and 3:1 is a second-order resonance, these period ratios cannot (without 
additional evidence from TTVs) be considered evidence for intervening planets.  In contrast, 16/9, 9/4 and 8/3 are fifth-order or 
higher, suggesting that they are not important resonances themselves and increasing the likelihood that there is an intermediate planet.  At third-order, a period ratio of 4 corresponds to a resonance that is likely to be weak, but cannot be as easily 
dismissed.

Several \ik multi-planet systems have observed planets with period ratios that are quite close to small integer ratios, but with the 
integers differing by more than two, suggesting associations with high-order resonances that would be quite weak unless the 
eccentricity of one or both bodies is large.  However, if one or two additional, as yet unseen, planet(s) orbits at an appropriate 
intermediate position, then the system could be connected via much stronger first-order resonances, as is KOI-730 (Section 
\ref{sec:res730}). We consider here two specific systems in some detail, then list several other KOIs that have interesting period ratios, and conclude this section with a discussion of situations where the planets may be near but not in resonance. 

The  observed period ratio of the pair of planets orbiting  KOI-657 is ${\cal P} = 4.00127$ (Table \ref{tab:two}).  This ratio is 
within one part in 3000 of the 4:1 resonance. There are only 12 neighboring planets with $3.5 < {\cal P} <  4.5$, so the {\it a 
priori} probability that any of these period ratios would be at least this close to 4 is $< 3\%$. But this resonance is third 
order, and thus it would require a large eccentricity to have significant strength.  In contrast, two 2:1 resonances with an 
unseen middle planet would be far stronger; the Hill separation $\Delta$ between the observed pair is 42.5, so there is
plenty of room for an intermediate planet, even if it is significantly more massive than the two that are observed.  Note that an 
alternative chain consisting of two intermediate planets in 2:1, 3:2 and 4:3 resonances, or three intermediate planets involving 
two 3:2 commensurabilities plus  two 4:3 commensurabilities, could also provide linkage via first-order resonances 

An analogous system is KOI-812, with the inner 2 candidates having a period ratio 
of 6.005814 (Table \ref{tab:three}). There are only 6 neighboring planets with $5.5 < {\cal P} <  6.5$, so the {\it a priori} 
probability that any of these period ratios would be at least this close to 6 is $\sim 7\%$.  The nearby 6:1 resonance is very 
weak, but a 3:1 and 2:1 with two intermediate planets would be significantly stronger, and two 
2:1 resonances plus a 3:2 resonance, requiring three intermediate planets, would be stronger still. The nominal $\Delta$ between 
these two observed candidates is 39, so there is
enough room for two or three intermediate planets of mass comparable to those expected for the observed candidates.

The candidate systems  KOI-82, KOI-117, KOI-124, KOI-313, KOI-961 and KOI-1203 also have pairs of planets near products of small integer ratios.  But these systems do not present as strong cases for undetected resonant candidates as the two systems mentioned in the title of this subsection because they have one or more of the following complicating factors: (i) At least one of the intermediate resonances would need to be the 5:4 and/or of second order (at most one 5:4 pairing exists among observed pairs, described above in Subsection \ref{sec:res191}, and it involves a giant planet and the period ratio deviates significantly from the ratio of the resonance integers). (ii) The estimated Hill separation, $\Delta$, required if the intermediate planet existed would be small. (iii) There is a non-resonant planet candidate within the suggested resonant planet chain. (iv) The observed period ratio deviates from the small integer ratio by a relatively large amount.

As discussed in Subsection \ref{secFreqRes}, much of the excess of observed near-resonant pairings involves period ratios of order 1\% larger than small integer ratios.  A search for chains with this magnitude of deviation from small integer ratios would require such a broad range of allowed period ratios that several pairs would likely  be identified even if no physical process was favoring such ratios.  However, such identifications may prove useful in searching for periods of unobserved  planets that might cause TTVs in observed candidates.

\section{Coplanarity of Planetary Systems} \label{sec:coplanar}
 
The orbits of the planets in our Solar System lie close to the same plane. The mean inclination of planetary orbits to the invariable plane of the 
Solar System is $\bar{i} \approx 2^{\circ}$; if Mercury is excluded, then $\bar{i} \approx 1.5^{\circ}$ and the set of inclinations is similar to a 
Rayleigh distribution of width $\sigma_i \approx 1^{\circ}$.  The strong level of co-planarity has been recognized as an important constraint on models 
of planet formation for over 250 years \citep{Kant:1755, Laplace:1796}, and our current understanding is that growth within a dissipative 
protoplanetary disk generally yields circular orbits and low relative inclinations \citep[e.g.,][]{Safronov:1969, Lissauer:1993}. The interplay between 
interactions with the protoplanetary disk, forcing by distant perturbers, and planet-planet scattering, leaves a fingerprint in the coupled 
eccentricity-inclination distribution of exoplanetary systems. High eccentricities are typical for RV-detected giant exoplanets whose orbital periods 
exceed one week. Some mechanisms that could produce these eccentric orbits would increase inclinations as well; planet-planet scattering is an 
example of such a process \citep[e.g.,][]{Chatterjee:2008, Juric:2008}. The inclination distribution of typical planetary systems provides an important 
test for planet formation theories.

 The inclination distribution of exoplanets is a fundamental aspect of planetary system dynamics. Yet neither transit observations, nor any other 
technique, have directly measured the true mutual inclination between planets observed in multiple systems, except in unusually fortuitous 
circumstances (e.g., \citealt{Wolszczan:2008,Correia:2010}). This continues to be true in multi-transiting systems; even though the inclinations to the 
line of sight of all transiting planets must be small, the orbits could be rotated around the line of sight and mutually inclined to one another. 
Indirect constraints, however, can be obtained for systems with multiple transiting planets that make such systems the best probe of mutual 
inclinations around main sequence stars. For example, the lack of transit duration variations of the planets of Kepler-9 \citep{Holman:2010a} and 
Kepler-11 \citep{Lissauer:2011} already place interesting limits on mutual inclinations to be $\lesssim$ 10$^{\circ}$, while the inclinations to the 
line-of-sight of the confirmed planet Kepler-10b and its recently-validated companion Kepler-10c require a mutual inclination $\gtrsim$ 5$^{\circ}$ 
\citep{Batalha:2011,Fressin:2011}. The lack of transit timing variations can also be used to constrain mutual inclinations \citep{Bakos:2009}. 
Occasionally it will be possible to measure mutual inclinations from exoplanet mutual events (Ragozzine et al., in prep.) or measurements of the 
Rossiter-McLaughlin effect of planets in the same system or by other means \citep{Ragozzine:2010}.

\begin{figure}
\begin{center}
\includegraphics[width=0.5\textwidth]{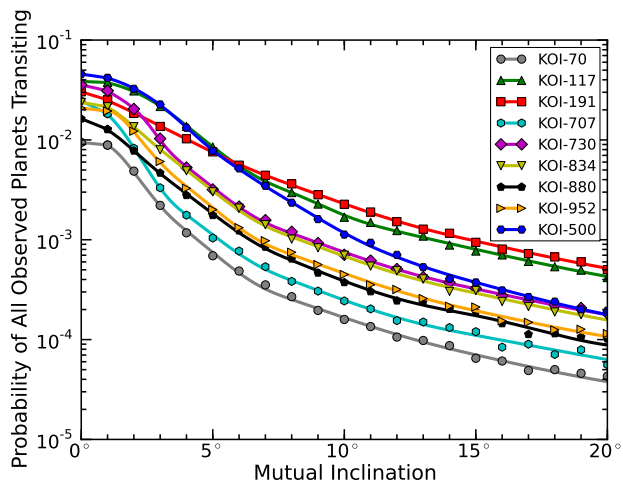}

\caption{Multiple transit probability for candidate four and five planet systems as a function of mutual inclination from Monte Carlo simulation. The 
curves correspond to individual systems as identified in the upper right. For each increment in mutual inclination, the multiple transit probability 
was computed by assigning a sky-frame inclination to each planet candidate in the system determined by the given mutual inclination with respect to an 
isotropic reference plane (i.e., a random observer's line of sight), and a random nodal angle \citep{Ragozzine:2010}. All orbits were assumed to be 
circular. Stellar sizes and masses were taken from Table 1 of B11.  A transit was defined as the center of the planet passing over any part of the 
star's disk. At low mutual inclination, where the planets are nearly coplanar, the probability of all the planets transiting is given by the geometric 
transit probability of the outermost planet. As the mutual inclination increases, the probability quickly decays, as detecting all the planets then 
requires a fortuitous alignment of the observer's line of sight and the orbital node of each planet.}
\label{fig:iprob}
\end{center}
\end{figure}

Given the large number of multi-candidate systems identified in B11, we can approach the question of the inclination 
distribution of the population statistically. The more mutually inclined a given pair of planets is, the smaller the probability that 
multiple planets transit \citep[Figure \ref{fig:iprob}; see also][]{Ragozzine:2010}, and thus the smaller the fraction of multi-planet systems that we would expect to see. 
Therefore, an investigation of coplanarity requires addressing the number of planets expected per planetary system. Of course, the 
question of planetary multiplicity is also inherently interesting to investigate. Here we provide a self-consistent 
estimate of the multiplicity and coplanarity of typical systems using a statistical approach involving Monte Carlo generation of simulated planetary 
systems. A similar study, undertaken contemporaneously with the second phase of our work, was recently presented by \citet{Tremaine:2011}, who take an analytical approach that is in some ways more general. 

\subsection{Simulated Population Model}

\subsubsection{Model Background}

The number of observed transiting planets per star is affected by several observational biases. The two most notable are that the probability of 
transiting decreases with increasing orbital period and that for a given size planet transiting a given star, the duty cycle (and thus likelihood of 
detection) is a decreasing function of orbital period, because the fraction of time that a planet spends in transit is $\propto P^{-2/3}$. With the 
current observations, we have little to no insight on the non-transiting planets in these systems, though eventually investigations of transit timing 
and duration variations (or the lack thereof) will be able to place limits on such planets \citep{Steffen:2010, Ford:2011}.  If we knew about the 
presence of every planet, transiting or not, then the problem of determining the typical multiplicity would be much easier; in this sense 
well-characterized radial velocity surveys have the advantage that they are much less sensitive to the inclination dispersion.

A useful way to address the problem of non-detections is to create a forward model of transiting planet detections based on a minimal number of 
assumptions and a relatively small number of tunable parameters and then to compare the output of this model with the properties of the observed 
systems. The model described here uses \ikt's observed frequency of planet multiplicity to estimate typical values for the true multiplicity of 
planetary systems as well as the mutual inclination between planets in the same system. After choosing a distribution for the number of planets 
assigned to a random \ik star (characterized by the parameter $N_p$) and their relative inclinations (characterized by $\sigma_i$), the model computes 
how many of these planets would be detected, requiring a geometrical alignment that leads to transit and a signal-to-noise ratio (SNR) large enough to 
be detected. This process is done for a sufficiently large number of simulated planetary systems (we use $10^6$ per simulation), and then the results are 
compared to the observed frequency of multi-transiting systems. Although imperfections in the model and both systematic and statistical uncertainties in the observations only allow an approximate answer, 
\ik data enable the first serious attempt at using observations to measure the typical values of $N_p$ (number of planets in a system) and $\sigma_i$ 
(characteristic inclination dispersion) for a sample consisting of a large number of planetary systems.

To isolate the distributions of multiplicity and inclination, other aspects of the simulated systems (such as radius and period distributions) must be 
consistent with the true underlying distributions. This requires some interpretation of the observed distribution and the identification of potential 
biases. Here, we hold the period and radius distributions fixed and pre-select them in a way that the output of the simulation will closely match the 
observed distribution. Our techniques for finding the debiased distributions differ from those used in \cite{Howard:2011} and \cite{Youdin:2011}, but 
is sufficient for the results discussed herein.

A good example of our methodology is the choice of the period distribution. The differences between the observed and true underlying period 
distributions are determined primarily by the geometric bias of observing planets in transit, which scales as $a^{-1} \propto P^{-2/3}$. As discussed 
in B11 and confirmed by our investigations, after correcting for the geometric bias and discarding very short periods ($P < 3$ days) and large planets 
($R_p > 6$ R$_\oplus$), the remaining \ik candidates are generally well described by a uniform distribution in $\log P$. The largest departure of the 
observed sample of $P > 3$ days, $R_p < 6$ R$_\oplus$ candidates from a uniform distribution in $\log P$ is an excess of observed planets near $\sim$20 
day orbital periods (semi-major axes near 0.1 AU), as discussed in B11. We consider a two-component period distribution: one component uniformly 
distributed in $\log P$ and another represented in a histogram as a Gaussian in $P$. The parameters of the Gaussian and the relative ratio of these two 
components were minimized by taking the resulting period distributions and comparing them to the observed distribution with a KS-test. The best fit was 
a Gaussian centered at 6.5 days with a width of 11.3 days (rejecting \emph{without replacement} periods below 3 days) where the fraction of periods 
drawn from a uniform distribution in $\log P$ was 27\% and that from the Gaussian in $P$ was 73\%. These values were determined by maximizing the similarity of the smooth two-component period distribution to the observed periods as measured by a KS-test. This two-component distribution gave a KS-statistic of 0.0289; for our sample sizes, one would expect a KS-statistic this large or larger 61\%  of the time, even if the two samples were drawn from the same distribution, i.e., the analytical period distribution is a good and smooth approximation to the observed period distribution for planets between 3 and 125 days and radii under 6 R$_\oplus$, which was the sole purpose of its construction. This analytic fit to the observed 
distribution of transiting planets is then debiased for the geometrical probability of alignment by assigning each period a weight proportional to 
$P^{2/3}$. This debiased analytical distribution worked well in our population simulations, but it should be noted that this distribution assumes that 
the probability of each planet transiting is independent of the multiplicity, which is not strictly true in that systems with higher multiplicity must 
accommodate more planets and tend to have slightly longer periods.

\begin{figure} 
\includegraphics[width=0.5\textwidth,angle=0]{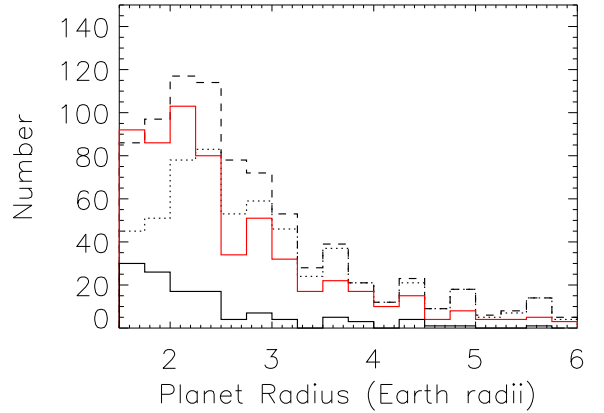}

\caption{ Comparison of the debiased radius distribution used in the simulated population (solid red line)
to the observed sample. The dashed histogram shows the radius distribution of all observed planets between 3 and 125 days with no limitations. The 
dotted histogram shows the subset of \ik planets with SNR $>$ 20 and periods between 3 and 125 days. This is the starting point for the debiasing 
process, which increases the relative proportion of smaller planets that cannot be detected around every star with existing candidates. The 
black solid histogram shows planets that would still be detected with SNR $>$ 20 even if they had radii of 1.5 Earth radii. The planets in the dotted 
histogram are weighted by SNR divided by the radius squared and the debiased distribution (solid red histogram) drawn from this weighted distribution. 
Here we show a random subset of the debiased radius distribution with the same number of planets as the dotted histogram to show how the relative 
weighting favors smaller planets.}

\label{fig:raddist}
\end{figure}

Estimating the true radius distribution of \ik planets also requires a debiasing process. A simple histogram (dashed in Figure \ref{fig:raddist}) of 
planet radii, $R_p$, from B11 shows that the majority of detected planets have radii around $\sim$2 Earth radii. Below $\sim$1.5--2 R$_\oplus$, the 
number of planets decreases significantly, which is at least partly due to small planets escaping detection due to low SNR. To correct for this, we 
restrict our analysis to planets with radii between 1.5 and 6 Earth radii and SNR greater than 20 (shown in the dotted histogram in Figure 
\ref{fig:raddist}). Each of these planets is given a weight of the observed SNR divided by the radius squared; this weight represents the SNR debiased 
based on planet size. Note that the weight was not assigned based on planet-star radius ratio, which tends to give extra weight to planets around small stars relative to larger stars. A planet radius distribution is then taken by randomly drawing from the weighted 
observed distribution. This distribution is shown by the red histogram in Figure \ref{fig:raddist}, which represents the underlying radius distribution 
and forms the radius distribution used by the simulated population.

There will be transiting planets drawn from this distribution that would not have been detectable by \ik in Quarters 0-2. An inspection of the B11 SNR ratios suggests that observational 
incompleteness sets in below an SNR of $\sim$20 (which is calculated from all transits of Quarters 0-5 binned together). In a trade-off between completeness and increasing the number of planets for statistical studies, 
a less conservative SNR cutoff of 16 is taken as the limit for a planet to be ``detectable.'' This SNR requirement will be applied to both the observed \ik population and the simulated populations. We did 
not consider the B11 vetting flag in our choice of acceptable systems as a large fraction of the multiples have not yet been investigated in detail. 

In summary, we are modeling the subset of \ik candidates that meet the following criteria (based on the properties as given in B11):
\begin{itemize}
\item The planet is detected in Quarter 0, 1, or 2; 
\item the orbital period is between 3 and 125 days;
\item the radius is between 1.5 and 6 Earth radii; and
\item the SNR (as listed in B11 for Q0-5) is at least 16.
\end{itemize}
Applying these cuts to \ikt's observed systems from B11 results in 479, 71, 20, 1, and 1 systems with 1, 2, 3, 4, and 5 acceptable transiting 
planets, respectively (Table \ref{tab:simpop}). Throughout this section, we will be referring only to this selected sub-population of qualifying planet candidates. These 572 qualifying systems are likely to include a small fraction of  false positives 
(primarily among the single-candidate systems) and our 
simulation does not attempt to account for this. Note that the sixth (outermost known) planet of Kepler-11, planet g, had a transit that fell into the 
data gap between Quarters 1 and 2, so this system counts as a quintuple for comparison to the simulated population.

\begin{figure}
\includegraphics[width=0.5\textwidth,angle=0]{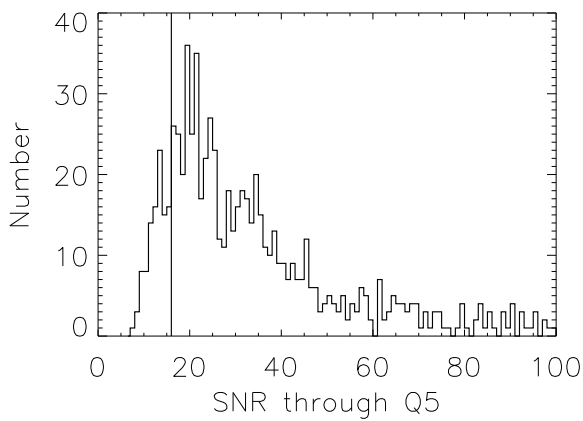}
\caption{SNR distribution of all candidates with periods from 3 to 125 days and radii between 1.5 and 6 Earth radii, as taken from Table 2 of B11. The departure 
from the trend of increasing number of planets at smaller SNR is due to observational incompleteness, which appears to be significant for SNR 
$\lesssim$~20. As a trade-off between completeness and increasing the number of planets for statistical studies, an SNR cutoff of 16 (vertical line) is 
taken as the completeness limit for the analysis of Section \ref{sec:coplanar}. Not shown are 48 candidates with a SNR greater than 100.}
\label{fig:snrhist}
\end{figure}

\subsubsection{Model Description} \label{sec:simpopmodel}

The model starts by taking the full stellar population from the \nall ~Q2 target stars in the exoplanet program (the list of which 
is available from MAST), with the assumed masses and radii from the \ik Input Catalog (KIC; \citealt{Latham:2005,Brown:2011}). Stars with radii larger than 10 $R_{\odot}$ and those with no KIC estimate of $T_{eff}$ or $\log g$ are removed, as are those which did not have a measured TMCDPP (the temporal median of the combined differential photometric precision) value, resulting in a list of 153599 stars. The distribution of stellar parameters for this subset is for our purposes statistically indistinguishable from the distribution including all \ik stars or the population used by B11.  

Under the constraints described above, we want to create simulated planetary populations to compare to the \ik observations. 
In particular, for each simulated population, we will compare the number of systems with $j$ detectable transiting planets to the observed distribution from \ikt. The two main parameters describing the simulated population is the average number of planets per star, $N_p$, 
and the inclination dispersion width, $\sigma_i$. 

The first parameter of the simulated population is the true multiplicity, $N_p$. Our goal was to employ a minimal number of parameters, so we focus on distributions of planetary multiplicities that can be characterized by a single value; an alternative approach of fitting an arbitrary distribution of planetary multiplicities is presented by \cite{Tremaine:2011}. Three distributions were considered 
for the number of planets: a ``uniform'' distribution, a Poisson distribution, and an exponential distribution. In the exponential 
distribution, the probability of having $j+1$ planets is a factor of $\alpha$ smaller than the probability of having $j$ planets, with 
$\alpha$ a free parameter. This model had a very poor fit to the data (anything that matched the triples to doubles ratio produced far too many systems 
with higher multiplicities), for any value of $\alpha$ or $\sigma_i$, so we do not consider it further.

We use $N_p$ to refer generally to the average number of planets per star and $N_{p,U}$ if these planets are distributed uniformly and $N_{p,\lambda}$ if
they are distributed according to a Poisson distribution. The ``uniform'' distribution parameter, $N_{p,U}$, is the mean number of planets per system, which is a fixed number of planets per star if $N_{p,U}$ is 
an integer or an appropriately proportioned mix of the two surrounding integers if $N_{p,U}$ is not an integer. For example, if $N_{p,U}=3.25$, 75\% of 
systems would be assigned 3 planets and 25\% would be assigned 4 planets. The Poisson distribution parameter $N_{p,\lambda}$ assigns the number of 
planets per system based on a Poisson distribution with mean $N_{p,\lambda}$. The frequency of stars with planetary systems is considered separately 
(see below), so we require the number of planets to be non-zero; for $N_{p,\lambda} \lesssim 2$, the distribution differs somewhat from a true Poisson 
distribution and $N_{p,\lambda}$ can be less than the average number of planets for the ensemble. These distributions were chosen because they can be represented by a single tunable parameter and are reasonable approximations to the 
true expected multiplicity distribution for the kinds of planets being considered.

The second parameter of the simulated population, $\sigma_i$, is the dispersion of the inclinations with respect to a reference plane. Planetary inclinations with respect to the reference plane in any given system were drawn from a Rayleigh distribution. The Rayleigh distribution is the appropriate choice for randomly distributed inclination angles \citep[see, e.g.,][]{Fabrycky:2009b} and requires a single value to define the distribution. The Rayleigh distribution is characterized by its width, $\sigma_i$; the mean value of the Rayleigh distribution with this width is $\sigma_i\sqrt{\pi/2}$. Note that this is equivalent to the true mutual inclinations between planets in these systems following a Rayleigh distribution with width $\sigma_i\sqrt{2}$.

Two different inclination distributions were considered, which we will call unimodal ($\sigma_{i,U}$) and Rayleigh of Rayleighs ($\sigma_{i,\cal{R}}$). In the unimodal distribution, the value for the Rayleigh parameter was taken to be the same for all systems in the simulated population, and we denote this value as $\sigma_{i,U}$.  In reality, the typical inclination dispersion for planetary systems will vary based on the number and masses of planets, the influence of the protoplanetary disk, and other mechanisms. Thus in a second set of simulations, the value of $\sigma_i$ for each system was itself  drawn from a Rayleigh distribution, so that the distribution of the individual simulated planets' inclinations relative to their reference planes is a Rayleigh of Rayleighs. We denote  the width of the Rayleigh distribution for the simulated population from which the Rayleigh parameter for each individual system was drawn by $\sigma_{i,\cal{R}}$. For small values of $\sigma_{i,\cal{R}}$, most systems are nearly coplanar (with a tail of a few systems with significant width); for larger values of $\sigma_{i,\cal{R}}$, a small number of systems are nearly coplanar, most have inclination widths near $\sigma_{i,\cal{R}}$, and a some have quite large inclinations. 

In all cases, nodal longitudes are assigned randomly. We assume zero eccentricities as, statistically, the observed multiplicity rates will not 
depend on eccentricities, assuming that the eccentricities and inclinations are not pathologically correlated. Note that one possible covariance is in systems where multiple planets are eccentric and the apses tend to be aligned or anti-aligned. Apsidal alignment occurs in the presence of dissipation and this could lead to an increased correlation between the probability of two planets transiting. This is a second-order effect, unless the eccentricities are typically significant and non-randomly aligned, and we neglected this effect in our calculations. 

After a specific non-zero number of planets is assigned to each star, the orbital and physical properties of these planets are assigned. First, the 
orbital period is assigned from the debiased analytical population described above. Each 
planetary radius is assigned based on the \ik observed radii independently of orbital period, as described above.
As almost all of the observed multiple planet systems are dynamically stable on long timescales (Section \ref{sec:stability}), we 
impose a proximity constraint that rejects simulated multiples that are too closely packed. This requires assigning a mass to each planet as well. We 
follow the mass-radius relation used above (Equation \ref{eq:mr}), i.e., $M_p=R_p^{2.06}$ with masses and radii in units of Earth masses and radii. The 
mutual Hill separation between each pair of planets, $\Delta$, can then be calculated as described in Equations (\ref{eq:rhill}) and (\ref{eq:delta}).  Although the period ratios in observed multiple systems deviate from a random distribution due to clustering near resonance (Section \ref{sec:res}), this is a minor effect compared to the overall geometric bias and the stability constraint that we impose. 

In practice, the systems are built one planet at a time until all the planets assigned to this star (based on $N_p$) are selected. The 
first planet is assigned its period, radius, and corresponding mass. Starting with the second planet, a new period, radius, and mass 
are assigned and $\Delta$ calculated. If any two planets have $\Delta < 3.46$ or if $\Delta_{\rm i} + \Delta_{\rm o} < 18$ for 
any three consecutive planets, then the planet is rejected on stability grounds and the process repeats until the appropriate number of planets have 
been assigned. Very rarely, over the wide value of $N_{p,\lambda}$ that we are considering, large numbers of planets (more than 12) are assigned to a particular star that will not fit within the period range and stability criteria; in this case, as many planets as can fit in 1000 tries are given to this star.

We can gain additional insight by considering 
systems from radial velocity, where even relatively large inclination dispersions 
do not significantly affect detectability. Using the assumed mass-radius relation 
(Equation \ref{eq:mr}) and accounting for the average random inclination to the line of 
sight, 
the size range we are considering corresponds to minimum masses of $\sim 1.8 - 
31.5$ M$_\oplus$. In the period range that we're considering, the largest number 
of planets in a radial velocity system that satisfy these constraints is 5 of the 
7 planets found around HD 10180 
\citep{Lovis:2011}. Furthermore, while adding more planets (up to $\sim$12) is possible theoretically in terms 
of the conservative dynamical stability requirements laid out above, values of $N_p \gtrsim 6$ are 
more likely to be long-term unstable in actual planetary systems. For this reason, we do not continue the calculations to distributions of systems with typical numbers of planets 
larger than $N_{p,U}=7$ and $N_{p,\lambda}=6$. 

At this point, the full orbital and physical characteristics of the planetary systems have been determined, and we can begin to assess observability. 
First, these planetary systems are rotated by a rotation matrix in a manner equivalent to choosing a random point on a sphere for the direction of the 
normal to the reference plane. In the new random orientation, the impact parameter of each planet is calculated. If the planets are transiting, the SNR 
of a single transit is calculated assuming a box-shaped transit with depth of $(R_p/R_\star)^2$ and duration calculated from Equation (15) in 
\cite{Kipping:2010}, corresponding to the time the center of planet crosses from one limb of the star to the other. A random epoch is assigned and the number of transits in a 127-day period (corresponding to the duration of Q1-Q2) is calculated. 
To account for the duty cycle of 92\%, each of these transits can be independently and randomly lost with 8\% probability. The total signal is then 
calculated as the signal for a single transit times the square root of the number of observed transits. This is compared to the estimated noise over 
the course of the duration of the transit, calculated from the TMCDPP values 
for the randomly chosen star from Quarter 2 (see \ik Data Release Notes available from MAST) to determine the simulated SNR. As described above, we require that our 
simulated planets have SNR of 9.2 in the simulated Quarters 1 and 2 (corresponding to a through-Q5 SNR of 16) in order to be detected. We find that for 
our various simulations, approximately 30--50\% of simulated transiting planets are ``missed'' due to insufficient SNR.

The major output of the model is the number of stars with $j$ detectable transiting planets, for all $j$. The next step is a statistical comparison 
between a variety of simulated populations (with different choices for $N_p$ and $\sigma_i$) with the observed distribution from \ikt. An assessment of 
whether a particular simulation is rejectable is done using an exact (non-parametric) test based on the multinomial probability of observing a distribution $O_j$, the 
number of systems with $j$ planets observed by \ikt, and $E_j$, the expected number of systems with multiplicity $j$ generated by scaling to the simulated 
population.

First, the two distributions are scaled so that the total values of $O$ and $E$ are the same, i.e., $\sum O_j = \sum E_j \equiv T$, where $T$ is the total number of \ik targets with one or more qualifying
observed planets ($T=572$ in the case of $1 \le j \le 6$). Since we will consider various comparisons to observed population (e.g., excluding 
singly-transiting systems ($O_1$)), we will describe the technique for the general case where we are comparing only planet counts where $j_{min} \le j \le j_{max}$. One could imagine then 
generating a very large set of Monte Carlo populations by randomly assigning planets to category $j$ with a probability $p_j \equiv E_j/ \sum E_j$. We 
have 
verified that the random probability of obtaining a distribution $x_j=[x_{j_{min}},x_{j_{min}+1},\ldots,x_{j_{max}}]$ in this Monte Carlo is equivalent 
to 
the expected 
multinomial distribution:
\begin{equation}
M_{x_j} \equiv \frac{T!}{x_{j_{min}}! \cdots x_{j_{max}}!} p_{j_{min}}^{x_{j_{min}}} \cdots p_{j_{max}}^{x_{j_{max}}}
\end{equation}
where $T = \sum_{j_{min}}^{j_{max}} x_j$. The multinomial distribution is the generalization of the binomial distribution when more than two outcomes are possible.

Due to the large number of possible populations, even the most probable distribution ($E_j$ itself) has a low probability of being exactly chosen. 
To test how likely it would be to draw the observed \ik observations from the distribution set by a particular simulated population, the significance, $S$, of the null hypothesis that $O_j$ is indistinguishable from $E_j$, is determined by 
calculating the sum of all the distributions of more extreme probabilities than observed \citep{Read:1988}:
\begin{equation}
S \equiv \sum_{x_j : M_{x_j} \le M_{O_j}} M_{x_j} . \label{eq:msig}
\end{equation}
If the observed and expected distributions are very similar, then there is substantial 
probability in values that 
are more ``extreme'' than the observed distribution and the significance is high ($S \approx 1$). If the observed and expected distributions are totally 
different, then 
the probability in values more extreme is very small, and the significance is very low ($S \approx 0$). The quantity $S$ is a measure of how good 
a match the two 
distributions are.\footnote{With some additional minor assumptions, this significance will be nearly equivalent to that given by the G-test (similar to 
the categorical $\chi^2$ test), which relies on calculating the G-statistic ($G \equiv \sum_{j} 2 O_j \ln (O_j/E_j)$) with no weight given to 
categories with no observed or expected systems.} Note that this method gives the most weight to categories with large numbers of objects, but no 
weight to categories where the simulated population produces 0 systems. For this reason, the simulated population contains a very large number of 
planetary systems, in order to assign at least minimal weight to categories that are rare but possible.

In practice, we calculate the logarithm (employing Stirling's approximation of the factorial) of the probabilities $M_{x_j}$ for a large scaled grid of values centered on $E_j$ moving outward until 99.5\% 
of the probability distribution is sampled; this is sufficient to calculate whether the hypothesis can be rejected with very high confidence. Note that the expected values from the simulated populations $E_j$ are only used in defining the probabilities $p_j$, so the scaling of $E_j$ to match the number of observed systems does not affect the calculation. We will refer to the fit between the \ik and simulated 
populations as ``adequate'' if the $S \ge 0.05$ as, in this case, we cannot reject with more than 95\% confidence the null hypothesis that the two 
populations are drawn from the same distribution. Even so, simulations with higher $S$ values are considered better fits to the observed distribution.

When the simulated population predicts 0 systems with $j$ transiting planets, yet such systems are observed, the multinomial statistic technically goes 
to 0, i.e., the simulated population is completely rejected. In reality, we are looking for a single population that fits the majority of the observed 
systems. In cases where the simulated population contains no quadruply or quintuply transiting systems ($E_4=E_5=0$), the two observed \ik systems 
with four or more planets can be considered outliers to the general model. To allow for this situation, the multinomial statistic is 
only evaluated over a range of $j$ values where the number of expected systems is non-zero (ignoring \ik observations outside this range, and thus limiting the test to fitting systems of lower multiplicity). 
Since triply transiting systems form a non-negligible fraction of the 
population, we do find it reasonable to reject outright simulated populations that produce no systems with three or more planets. 

In the populations where the number of planets per system is drawn from a Poisson distribution, the simulated population 
will have significant numbers of systems with $j>5$ transiting planets, where \ik observations saw no systems. In this case, the multinomial statistic can still calculate 
the probability that the simulated population matches the \ik observations. However, the calculation of the multinomial statistic increases 
significantly in difficulty with the number of separate $j$-planet bins in the distribution. For simplicity, we have chosen to lump all simulated systems with 6 or more planets into a single bin, i.e., comparing 
$O_{6+}=0$ to $E_{6+} \equiv E_6+E_7+\ldots$ in the multinomial statistic. Throughout, 
reference to $j=6$ implies this binning, i.e., we will not distinguish between $E_6$ and $E_{6+}$. Most of the weight is given to the categories with more planets, so the arbitrary choice of cutting the distribution off at 6 does not affect our conclusions. 

The simulated population assumes that all $10^6$ Monte Carlo stars have planetary systems. B11 report that the expected number of planets per star 
based 
on the \ik observations is about 0.34, though this describes a wider range of planets than we are considering here; for the limited size range we are 
considering, the B11 value is $\approx$ 0.2. Note that converting this to the fraction of stars with planetary systems (one or more planets) requires 
dividing by 
the typical multiplicity. 

Our investigation could address the fraction of stars with planetary systems, $f_p$, by adding a new free parameter and fitting to $O_j$ 
with $0 \le j \le 6$. We choose instead to compute $f_p$ by first fitting a scaled population of systems that all contain planets to $O_j$, $j>0$, and then 
calculating the scaled number of systems that had planets but did not transit, $E_0$. Since $\sum_{j=0}^6 E_j$ is the number of stars with planetary systems needed to match the number of planets observed by \ikt, we find $f_p=\sum_{j=0}^6 E_j /\nall$. That is, we can answer the 
question of the frequency of planetary systems after we find simulated populations that match the observed frequencies of systems where at least one 
transiting planet is detected.  Note that for all simulations that have a uniform number of planets per star that hosts planets, $N_{p,U} \ge 2$, so in these models the number of planetary systems is identical to the number of multi-planet systems.  When we fit only for observed numbers of multi-planet systems, the number of singly-transiting systems are underpredicted, and the additional single planet systems that would be needed to make up this deficit are not included in $f_p$. In contrast, when we use a non-zero Poisson distribution for the true number of planets per star hosting at least one planet, then some stars are assumed to host one planet, and these are included as planet hosts in computing $f_p$. 

\subsection{Results and Discussion}

\begin{figure}
\includegraphics[width=0.5\textwidth]{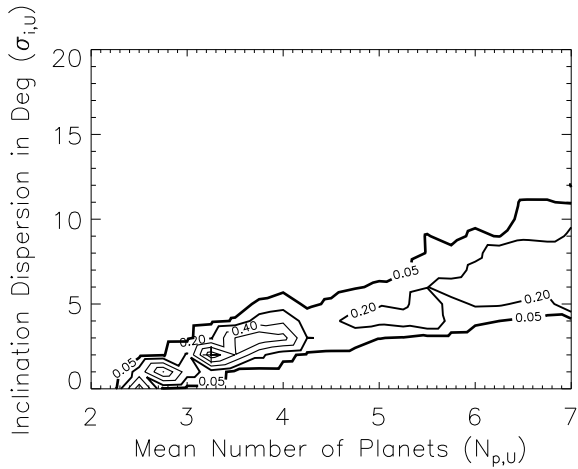}

\caption{Contour plot of the significance ($S_{2,6}$, Equation \ref{eq:msig}), i.e., the probability that a simulated population matches the \ik observed 
population of systems with 2 or more 
transiting planets with radii between 1.5 and 6 Earth radii, periods between 3 and 125 days, and with SNR of 9.2 or greater (see 
Section \ref{sec:coplanar} and Table \ref{tab:simpop}). The horizontal axis shows the average number of planets assigned to all stars in the ``uniform'' model and the vertical 
axis shows the inclination width as drawn from a unimodal Rayleigh distribution
$\sigma_{i,U}$ for the mutual inclinations in degrees. A plus sign marks the best-fit model  
described in the text and Table \ref{tab:simpop}. Contours showing $S_{2,6}$ equal to 0.05, 0.2, 0.4, 0.6, and 0.8 are shown. Models outside the thicker 0.05 contour would be rejected as unacceptable fits with an 95\% confidence level. 
Note the clear trend that,  in order to match the \ik observations, the simulated population must 
have a higher inclination dispersion if the number 
of planets per star is larger. For this multiplicity distribution, populations with mean inclinations greater than $\bar{i} 
\equiv \sigma_i 
\sqrt{\pi/2} \approx 5^{\circ}$ are adequate, but not as good as fits with smaller inclination dispersions. A contour plot considering the significance 
$S_{1,6}$ for the same parameters has lower values in general, and they are shifted such that at the same value of $N_{p,U}$ a higher inclination dispersion (by about $\sim$2$^{\circ}$) is needed.}
\label{fig:mspuni}
\end{figure}

\begin{figure}
\includegraphics[width=0.5\textwidth, angle=0]{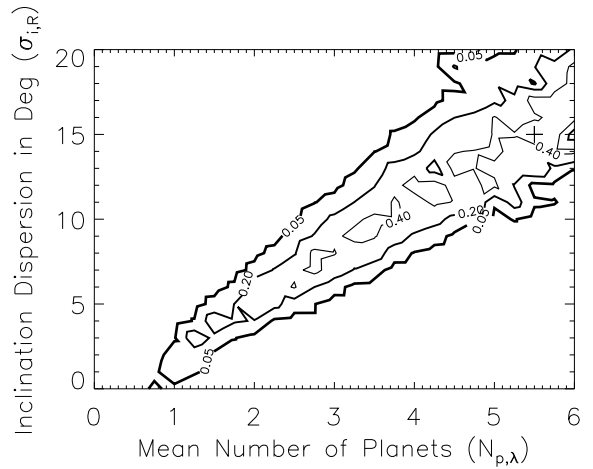}

\caption{Same as Figure \ref{fig:mspuni} but where the number of planets (in the specified range) is drawn from a Poisson distribution (without allowing zero 
planets) with width $N_{p,\lambda}$ and the inclination dispersion is given by a Rayleigh of Rayleighs with width $\sigma_{i,\cal{R}}$. In this figure, we show contours of $S_{1,6}$, i.e., these simulated populations were fit to all \ik transiting planet systems. These populations are worse (but still adequate) fits to the observed numbers than 
the model shown in Figure \ref{fig:mspuni}, and show the same trend of increased inclination dispersion required at higher multiplicity. The \ik observations alone cannot put a strong limit on the true multiplicity of planetary systems.}

\label{fig:msplam}
\end{figure}

The main results of our simulated populations are shown in Table \ref{tab:simpop}. We find significant differences in the best-fit simulated populations when fitting to $O_j$ over the entire range $1 \le j \le 6$ compared to fits that do not attempt to match the 479 observed singly-transiting systems. We distinguish between the significance with and without the singly-transiting systems by defining $S_{1,6}$ to be the significance when comparing $O_j$ and $E_j$ for $1 \le j \le 6$ (i.e., $j_{min} = 1$ in Equation \ref{eq:msig}) and $S_{2,6}$ to be the significance for $2 \le j \le 6$. Both of these significances are listed in Table \ref{tab:simpop}. That table lists the best fit from each of the eight possible combinations using $N_{p,U}$ or $N_{p,\lambda}$, $\sigma_{i,U}$ or $\sigma_{i,\cal{R}}$, and $j_{min}  = 1$ or $j_{min} = 2$. Two of the extreme simulations (for $N_{p,\lambda}$ and $\sigma_{i,\cal{R}}$) that are very poor fits are also given to show how the \ik observations are able to rule out large inclinations with small numbers of planets and coplanar systems with large numbers of planets.   Our simulations spanned a much larger range than represented in the table, as can be seen in Figure \ref{fig:mspuni}, which shows a contour plot of $S_{2,6}$ that includes the best fit to multiple planet systems ($N_{p,U}=3.25$, $\sigma_{i,U}=2^{\circ}$, $S_{2,6}=0.94$), and Figure \ref{fig:msplam}, which shows a contour plot of $S_{1,6}$ that includes the best fit including singly-transiting systems ($N_{p,\lambda}=5.5$, $\sigma_{i,\cal{R}}=15^{\circ}$, $S_{1,6} = 0.46$).  Both of these figures demonstrate the trend that more planets per star that has planets implies that a larger 
inclination dispersion is required to match the \ik observations. 

\begin{deluxetable*}{lccrrrrrrrrrrr}
\tablewidth{0pc}
\tablecaption{ \label{tab:simpop}  }
\tablehead{
\colhead{Name} \strut & \colhead {$N_p$} & \colhead{$\sigma_i$} & \multicolumn{7}{c}{Number of systems with $j$ transiting planets ($E_j$)} &
\multicolumn{2}{c}{Significance ($S$)} & \colhead{$<N>$} & \colhead{$f_p$} \\
\colhead{} \strut & \colhead{} & \colhead {} & \colhead{0} & \colhead{1} & \colhead{2} & \colhead{3} & \colhead{4} & \colhead{5} &
\colhead{6+} & \colhead{$S_{1,6}$} & \colhead{$S_{2,6}$} & \colhead{} & \colhead{}
}
\startdata
\hline
All \ik \strut & \nodata & \nodata & 159210 & 791 & 115 & 45 & 8 & 1 & 1 & \nodata & \nodata & $\sim$0.34 & \nodata  \\
Selected \ik ($O_j$) \strut & \nodata & \nodata & 159719 & 479 & 71 & 20 & 1 & 1 & 0 & \nodata & \nodata & $\sim$0.2 & \nodata \strut \\
\hline
Simulated & $N_{p,U}=5.75$ \strut & $\sigma_{i,U} = 10^{\circ}$ & (4694.5) & 469.9 & 88.4 & 12.2 & 1.3 & 0.1 & 0.0 & 0.01 & ($<$0.01) & 0.189 & 0.033 
\\
Simulated & $N_{p,U}=3.25$ \strut & $\sigma_{i,U} = 2^{\circ}$ & (5258.3) & (174.2) & 72.3 & 18.9 & 1.8 & 0 & 0 & ($<$0.01) & 0.94 & 0.112 & 0.034 \\
Simulated & $N_{p,\lambda}=4.0$ \strut  & $\sigma_{i,U} = 9^{\circ}$ & (7281.9) & 465.7 & 88.6 & 15.3 & 2.1 & 0.3 & 0.0 & 0.11 & (0.11) & 0.196 & 0.049 
\\
Simulated & $N_{p,\lambda}=1.75$ \strut & $\sigma_{i,U} = 1^{\circ}$ & (9481.2) & (229.4) & 68.9 & 18.7 & 4.4 & 1.0 & 0.0 & ($<$0.01) & 0.51 & 0.107 & 
0.061\\
Simulated & $N_{p,U}=5.75$ \strut & $\sigma_{i,\cal{R}} = 10^{\circ}$ & (4826.9) & 472.7 & 81.9 & 14.2 & 2.6 & 0.5 & 0.06 & 0.25 & (0.17) & 0.194 & 
0.034\\
Simulated & $N_{p,U}=3.25$ &\strut  $\sigma_{i,\cal{R}} = 2^{\circ}$ & (5548.2) & (199.1) & 72.2 & 18.8 & 2.0 & 0 & 0 & ($<$0.01) & 0.86 & 0.119 & 
0.036\\
Simulated & $N_{p,\lambda}=5.5$ & $\sigma_{i,\cal{R}} = 15^{\circ}$ & (5507.3) & 473.7 & 77.9 & 15.5 & 3.4 & 1.0 & 0.46 & 0.56 & (0.43) & 0.209 & 
0.038 \\
Simulated & $N_{p,\lambda}=2.25$ \strut & $\sigma_{i,\cal{R}} = 3^{\circ}$ & (8892.9) & (286.2) & 69.3 & 18.0 & 4.3 & 1.1 & 0.3 & ($<$0.01) & 0.52 & 
0.130 & 
0.058 \\
\hline
Simulated & $N_{p,\lambda}=0.5$ \strut & $\sigma_{i,\cal{R}} = 10^{\circ}$ & (76302.0) & (1960.7) & 87.7 & 5.0 & 0.3 & 0.0 & 0.0 & ($<$0.01) & $<$0.01 
& 0.622 
& 0.489 \\
\bigskip
Simulated & $N_{p,\lambda}=4.5$ & $\sigma_{i,\cal{R}} = 0^{\circ}$ & (4134.3) & (84.8) & 43.1 & 24.0 & 14.1 & 7.9 & 7.0 & ($<$0.01) & $<$0.01 & 
0.121 
& 0.027 \\
\enddata
\tablecomments{ Summary of best-fit simulated populations from Section \ref{sec:coplanar}. The selected planets have radii between 1.5 and 6 Earth 
radii, periods 
between 3 and 125 days, SNR greater than 16, and show at least one transit in Quarter 0, 1, or 2. The second line shows the number of \ik systems that 
meet these requirements, and it was this distribution of multiplicities that was compared to the simulated populations. The second column shows whether 
the simulated population in question had a true planetary multiplicity drawn from a uniform distribution ($N_{p,U}$) with specified mean or a non-zero 
Poisson distribution ($N_{p,\lambda}$). The third column shows whether the Rayleigh parameter specifying the inclination dispersion from which the 
planets in a given system were drawn was the same for all systems ($\sigma_{i,U}$) or whether the value of this parameter was itself Rayleigh 
distributed ($\sigma_{i,\cal{R}}$), as well as the value of $\sigma_i$. See Section \ref{sec:simpopmodel} for details. The number of systems with 0, 1, 
2, 3, 4, 5, and 6 or more detectable transiting planets is shown, along with the significance $S$ (see Equation \ref{eq:msig}) of the multinomial 
statistic when compared to the \ik observations of $1 \le j \le 6$ planets ($S_{1,6}$) and $2 \le j \le 6$ planets ($S_{2,6}$). Higher values of $S$ 
are better fits, and populations with $S$ values less than 0.05 are not adequate since they can be rejected at the 95\% confidence level. The simulated 
population is shown scaled to the observed population (i.e., $\sum_{j_{min}}^{j_{max}} O_j = \sum_{j_{min}}^{j_{max}} E_j$) for the range of $j$ values 
implied by the $S$ column that is not in parentheses. Further, the numbers of those classes of systems that are not fit attempted to be fit by this 
group of simulations are shown in parentheses.  The best fit for each of the four possible combinations of $N_p$ and $\sigma_i$ is shown for both 
$S_{1,6}$ and $S_{2,6}$ above the line. Two poorly-fitting simulations are presented below the line. The column labeled $<N> \equiv \frac{N_p \sum_j 
E_j}{\nall}$ gives the mean number of planets per star for the observed and simulated populations and $f_p \equiv \frac{\sum_j E_j}{\nall}$, the 
fraction of stars with planetary systems \citep[see][]{Youdin:2011}. See the text for discussion. The $N_{p,U}$ and $\sigma_{i,U}$ values were tested 
over the range 2-7 and 0-20$^{\circ}$, respectively, as shown in Figure \ref{fig:mspuni}. The $N_{p,\lambda}$ and $\sigma_{i,\cal{R}}$ values were 
tested over the range 0.5-6 and 0-20$^{\circ}$, as shown in Figure \ref{fig:msplam}. The $N_{p,U}$ and $\sigma_{i,\cal{R}}$ values were tested over the 
range 2-6 and 0-10$^{\circ}$ and the $N_{p,\lambda}$ and $\sigma_{i,U}$ values were tested over the range 0.5-4.5 and 0-10$^{\circ}$, respectively.}
\end{deluxetable*}

When looking at the fits to the entire \ik sample, including singly-transiting systems, the observed 
distribution is fit by having large numbers of planets (in the specified period and radius range) in most systems and only 
observing a small fraction of them in transit due to the large inclination 
dispersion.  It is clear from Table \ref{tab:simpop} that the Rayleigh of Rayleighs inclination dispersion is a much better match than the unimodal inclination dispersion to the \ik sample including singly-transiting systems, presumably because it allows more multiples to have large enough inclinations that observing only a single planet is probable, while still matching the relatively high ratio of triples to doubles ($O_3/O_2 = 0.282$). A contour plot showing the significance $S_{1,6}$ as a function of $N_{p,\lambda}$ and $\sigma_{i,\cal{R}}$ is shown in Figure \ref{fig:msplam}. We do not include constraints from radial velocity surveys, but the higher multiplicities ($N_p \gtrsim 6$) would likely lead to many more tightly packed and high multiplicity systems than are known. 

Given the small number of degrees of freedom (5 multiplicities -- 1 scaling -- 2 parameters), it is not surprising that there are a large range of adequate fits. As expected, Figure \ref{fig:msplam} shows the degeneracy that increasing the number of planets per system requires the inclination 
dispersion to increase in order to provide a good match to the \ik data. Any population within a contour line is an acceptable fit in 
that it cannot be rejected at 95\% confidence that the simulated populations with these parameters are significantly different than the 
observed \ik systems. Regions within multiple contours provide even better fits. 

\begin{figure}
\includegraphics[width=0.5\textwidth,angle=0]{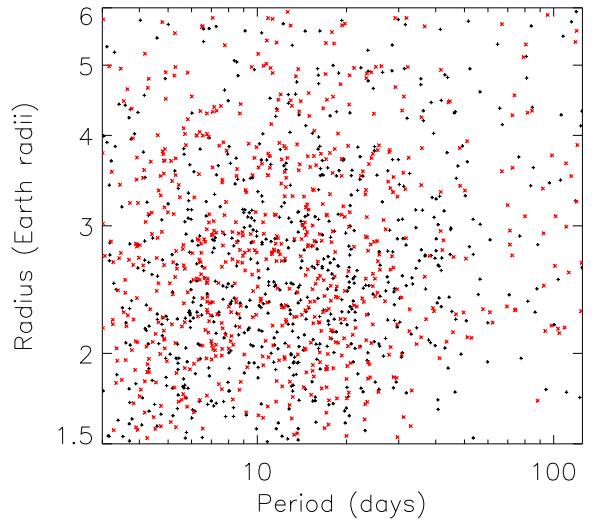}
\caption{
Comparison of the period-radius distribution of \ik planets to the best-fit simulated population described in Section \ref{sec:coplanar} and Table 
\ref{tab:simpop}. Black plus signs show the radius-period distribution 
of observed \ik planets that satisfy the SNR, and other criteria for inclusion of the Observed population that is to be compared to the simulated 
populations. The plot only shows periods from 3 to 125 days and radii between 1.5 and 6 Earth radii, which 
are the limits discussed in Section \ref{sec:coplanar}. Red $\times$'s show the simulated planets from our best-fit population ($N_{p,U}=3.25$ and 
$\sigma_{i,U}=2.0^{\circ}$), with the number of planets scaled to match the observed distribution. One and two-dimensional KS tests show that the observed 
and simulated distributions are similar. As discussed in \citet{Howard:2011} and \citet{Youdin:2011}, the first 126 days of \ik data is not complete 
below 2 Earth radii and the 
apparent drop off in planets between 1.5 and 2 Earth radii is due to incompleteness effects; the actual population probably continues to increase in 
number for smaller and smaller planets. 
} \label{fig:simpopperrad} 
\end{figure}

Not every star has planets in the size and period range that we are considering here; the Sun is an example of a star lacking such planets. The 
number of stars required to produce the observed population is given by summing over the $E_j$ columns in Table \ref{tab:simpop} and ranges from approximately 5000--10000. The mean number of planets per star, given as $<N> \equiv \frac{N_p \sum_j E_j}{\nall}$ in Table \ref{tab:simpop}, range from 0.182 -- 0.196 for simulations that best fit $1 \le j \le 6$. When including the required additional population of single planets to the simulated populations that best fit $2 \le j \le 6$, the mean number of planets per star is expected to be similar. These values are slightly lower than $\sim$0.20 planets per star estimated by B11 for the subset of planets we are considering here. This small discrepancy could be partly explained by completeness issues. We 
chose an SNR = 16 cutoff that included more planets at the cost of being somewhat incomplete, especially in the range 16 $<$ SNR $<$ 20. Correcting for 
this completeness would increase the number of simulated planetary systems needed to match the observations, bringing up our estimates to that 
calculated in B11.

\begin{figure}
\includegraphics[width=0.5\textwidth,angle=0]{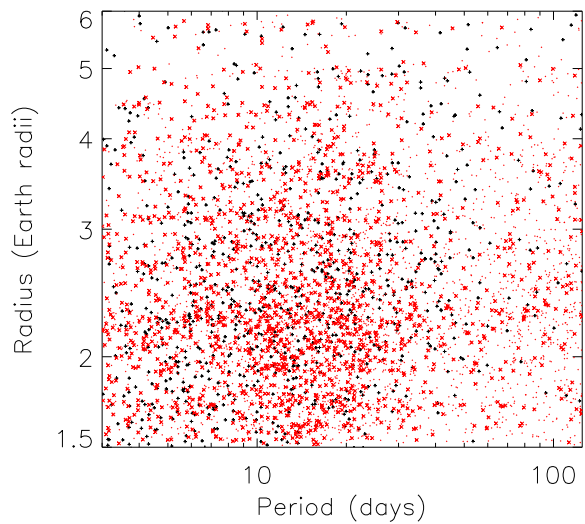}
\caption{
Similar to Figure \ref{fig:simpopperrad}, but including planets with low SNR. Black plus signs again show \ik planets in the same period and radius 
range, including now all detected planets irrespective of SNR. The red $\times$'s show planets in the scaled simulated population that were rejected as 
``undetectable'' at present, due to insufficient SNR, but which would reach a SNR $\ge$ 16 in an extended 6-year \ik Mission. Thus, the union of the 
``+'' and ``$\times$'' distributions represents an estimate of what a plot of \ik planets detected at SNR $>$ 16 using 6 years of data will look like. 
Red dots show simulated planets that were non-transiting in systems with at least one transiting planet. As expected for a relatively thin population, 
the number of non-transiting planets increases significantly at longer periods.  The distributions of these additional 
planets depends on the simulated population, but the results are qualitatively similar for other simulated populations that are adequate fits to the 
observed multiplicities (see Section \ref{sec:coplanar}). } 
\label{fig:simpopperradmore} 
\end{figure}

The simulated population models with unimodal inclination dispersions ($\sigma_{i,U}$) do much better when they do not attempt to match the number of observed singles. The difficulty in simultaneously matching singles and multis can be seen 
qualitatively by comparing the ratio of doubly-transiting systems to singly-transiting systems ($O_2/O_1 = 0.148$) to the ratio of 
triply-transiting systems to doubly-transiting systems ($O_3/O_2 = 0.282$). These two ratios are quite different, yet the probability 
of an additional planet transiting depends only on the semi-major axis ratio and the mutual inclination \citep{Ragozzine:2010,Tremaine:2011}, regardless of the 
number of planets. We have 
also found that the probability that a planet passed the SNR detection threshold is also more-or-less independent of multiplicity, i.e., the 
detectability can be estimated on a roughly planet-by-planet basis. (In detail, systems with high multiplicities tend to have planets with 
longer periods, which typically lowers the duty cycle, and therefore the summed SNR, somewhat.) The simulated populations have both the geometric and detectability probabilities 
decreasing at an approximately constant rate as additional planets are added, generally leading to ratios $E_{j+1}/E_j$ that are nearly constant.

Unlike in the majority of the simulated populations, the observed $O_{j+1}/O_j$ ratios are not similar and \ik may be seeing an ``excess'' of 
singly-transiting systems. Note that this cannot be attributed to the apparent fact that ``hot Jupiters'' appear to be 
singletons \citep{Latham:2011}, since we are considering objects that are smaller than 6 R$_\oplus$. Some of the excess 
singly-transiting systems could be false positives, but we find it unlikely that the majority of the shortcoming can be ascribed 
to bad candidates. An excess of singles and/or very high inclination multi-planet systems, in the form of a population distinct from the one providing the overwhelming majority of observed multi-transiting systems, is strongly suggested on theoretical grounds for long-period planets 
due to a distinction between systems that have had large dynamical scatterings in the past versus those systems that have remained 
relatively calm \citep[e.g.,][]{Levison:1998}, and it may be present in the short-period population we are modeling here.

As we are only attempting to match at most 6 multiplicity frequencies $O_j$, we cannot justify adding a full second population to our fits 
since this would require 4 highly degenerate parameters. Instead, we investigate the fits only to the observed population with two 
or more transiting planets (i.e., $2 \le j \le 5$) using the same method described above. Most of the constraint on this 
population is from the ratio of $O_3/O_2$. It is important to note that the comparison statistic $S$ is not modified when there are fewer degrees of freedom, so that the larger values of $S_{2,6}$ compared to $S_{1,6}$ may not imply better fits in a statistically significant way.

The best fit for multis alone ($S_{2,6}=0.94$) is given by  a population of systems with 
$N_{p,U}=3.25$ 
(i.e., 75\% of systems with 3 planets and 25\% of systems with 4 planets) and with a single Rayleigh inclination width of $\sigma_{i,U}=2^{\circ}$, 
implying a mean inclination of $\bar{i}\approx2.5^{\circ}$. This population produced an excellent match to the observations, though it obviously cannot explain systems with 5 or more planets (but see Section \ref{sec:K11}). This population produced only 
$E_1=174.2$ singly-transiting systems. The under-production of singly-transiting systems is a trend that is shared with all the best-fit distributions with $2 \le j \le 6$, suggesting that as many as two-thirds of the observed singles could derive from a different distribution 
than the simple population model that matches the higher multiplicities.  Figure \ref{fig:mspuni} gives a contour plot of the acceptability of simulated 
populations as a function of $N_{p,U}$ and $\sigma_{i,U}$ for $2 \le j \le 6$.

Figure \ref{fig:mspuni} also demonstrates the trend that more planets require a larger 
inclination dispersion to match the \ik observations. In this case, the observed 
distribution is fit by having large numbers of planets in most systems and only 
observing a small fraction of them in transit due to the larger inclination 
difference. In all the fits we are considering here, this tends to create a period distribution of the simulated 
``observed'' planets in multiple systems that is much more heavily weighted towards periods 
$\gtrsim$40 days than is the observed population. Furthermore, the orbital 
separation as measured by mutual Hill radii, in a plot similar to Figure 
\ref{fig:hill}, show systems with generally tighter distributions than in the \ik 
population, as this maximizes the probability that nearby planets both transit 
when mutual inclinations are high. Despite these shortcomings, we cannot reject 
these simulated populations with large inclination dispersions using the statistical tests that we have performed. We leave
for future work simulated populations that attempt to simultaneously fit the multiplicity, period, and
mutual Hill radii distributions and which include information of planetary multiplicity from radial velocity surveys.

\subsection{Conclusions of Coplanarity Study}

\ik is providing incredibly powerful insight into the structure of planetary systems. This simulated population model is the first to explore the multiplicity and inclination distribution of these planetary systems, with results summarized in Table \ref{tab:simpop}. \ik has elucidated a new population of planetary systems 
suspected from RV observations: \emph{a few percent of stars have multiple similar-sized, somewhat-coplanar, 1.5 -- 6 R$_\oplus$ planets with periods between 3 and 125 days}.  While the distribution of \ik candidates implies that few of these systems contain Jupiter-size planets, the true multiplicities including planets of any size (below the 1.5 R$_\oplus$ limit) will actually be higher than the numbers described here.

There is some evidence that the 
observed distribution may require more diversity than a single homogeneous population. To match the 
numbers of higher multiples observed by \ik requires a population with relatively low inclination dispersion; mean inclinations 
less than $\sim$ 10$^{\circ}$ are preferred. If all the systems are drawn from the same source population, then the mean inclinations could be much higher (though this is not necessary to obtain an adequate fit).  However, the high planetary multiplicities required to obtain adequate fits for high inclinations may not be consistent with radial velocity surveys \citep[see also,][]{Tremaine:2011}.   

As expected, our simulated populations suggest that the majority of the doubly-transiting systems and a substantial fraction of the singly-transiting systems are probably systems with multiple 
planets in the size-period intervals considered for this study (see Figure \ref{fig:simpopperradmore}). Systems with large values of $\Delta$ in Tables \ref{tab:two} and \ref{tab:three} are good candidates 
for systems where additional 
non-transiting planets or undetectably small planets may be missing in between observed planets. Similarly, systems with small values of $\Delta$ could easily have external planets 
that are not transiting (and possibly even non-transiting internal planets). This is consistent with the observation that $\sim$15\% of candidate planets show significant TTVs and that the frequency of 
candidates that show TTVs is independent of the number of candidates observed per system \citep{Ford:2011}. Our simulated populations are in general 
agreement with the corrected population models of B11, \citet{Howard:2011}, and \citet{Youdin:2011}, but suggest that planets may be less abundant than estimated from radial velocity surveys \citep{Howard:2010,Mayor:2011}.

Explaining the formation and evolution of this population of 
systems in the context of other types of apparently distinct systems (e.g., hot Jupiter systems and the Solar System) should be a major goal of 
planet formation theories.  Systems with multiple super-Earth-size and Neptune-size planets with periods less than 125 days at low relative 
inclinations suggest interactions with a protoplanetary disk that induced migration while damping inclinations \citep[e.g.,][]{Bitsch:2011}. Note that 
inclinations can be excited during disk migration when planets are captured in resonances \citep{Lee:2009}. The scattering process that produces the large eccentricities observed for giant planets, which 
would also produce large inclinations \citep{Chatterjee:2008,Juric:2008,Libert:2011}, but the distribution of transit durations of \ik planets does not show evidence for large eccentricities \citep[see 
also][]{Moorhead:2011}.

There are good prospects for improving our understanding of the planetary populations observed by \ikt. Besides continued observations 
that will increase the detectability of planets and find new candidates in multiple systems, a more detailed population simulator 
could be developed. For example, although the errors in the impact parameters of \ik planets are usually relatively large, impact parameters do 
contain information on relative inclinations \citep[e.g.,][]{Fressin:2011} that was not used in the study presented herein. 

In some systems, the detection of or upper limits to TTVs or Transit Duration Variations (TDVs) will constrain the unobserved component of the 
relative inclinations of planets \citep[e.g.,][]{Holman:2010a}.  Potential exoplanet mutual events in \ik multi-planet systems and their 
constraints on mutual inclinations will be studied in more detail by \cite{Ragozzine:2011}. Measurements of the 
Rossiter-McLaughlin effect for planets within multi-planet systems would constrain orbital inclinations to star's equator as well as the mutual 
inclination between planets \citep{Ragozzine:2010}. Each measurement of the true mutual inclination in individual 
systems helps fill in the picture for the typical inclination distribution of planetary systems. Unfortunately, measurements of true mutual inclinations will be observationally challenging for most \ik stars.

\section{How Rare are Planetary Systems Similar to Kepler-11?}\label{sec:K11}

The Kepler-11 (= KOI-157) planetary system has six transiting planets, whereas only one other \ik target has even as many as five planetary candidates 
identified to date.  Moreover, the period ratio of the two inner planets of Kepler-11 is only 1.264, which is the 3$^{rd}$ lowest ratio among the 238 
neighboring pairs of transiting {\it Kepler} exoplanets (Figure~\ref{fig:cumprat}).  The smallest period ratio, 1.038, is between two weak (vetting flag 3) candidates in 
the KOI-284 system; this system would be unstable for any reasonable planetary masses if indeed both of these candidates are actually planets and they orbit 
about the same star, so we suspect that these candidates do not represent two planets in orbit about the same star (Section~\ref{sec:stability}). The second smallest period ratio, 1.258, is for a pair of candidates in the KOI-191 system that 
stability considerations strongly suggest are locked in a 5:4 mean motion resonance that prevents close approaches (Subsection~\ref{sec:res191}).  Thus, 
Kepler-11b and Kepler-11c may well have the smallest period ratio of any non-resonant pair of planets in the entire candidate list.  Additionally, the 
five inner planets of Kepler-11 travel on orbits that lie quite close to one another, both in terms of period ratio and absolute distance. Kepler-11 
clearly does not possess a ``typical'' planetary system, but are systems of this type quite rare, or simply somewhat scarce?

Although its solitary nature makes determining a lower bound on planetary systems like Kepler-11 an ill-posed question ({\it Kepler} could have just 
gotten lucky), we can estimate the number of Kepler-11 systems that would not be seen in transit probabilistically. If all the Kepler-11 planets were 
coplanar (a slight non-coplanarity is required by the observed impact parameters), then the probability of seeing all six planets transit would be the same as 
seeing Kepler-11g transit, which is 1.2\%, suggesting a simple estimate of \nall ~$\times$ 0.012, or $\sim$ 1 per 2000 stars. However, the 
better-fitting population simulations described in Section \ref{sec:coplanar} rarely produced with systems of 5 or more planets (see Table \ref{tab:simpop}), raising the 
possibility that Kepler-11 
is not just a natural extension of \ik multiple systems.

The two candidate multi-planet systems most analogous to Kepler-11 are KOI-500 and KOI-707.  With 5 planet candidates, KOI-500 is second in abundance to 
Kepler-11, and four of these candidates are close to one another in terms of period ratio (Table~\ref{tab:fiveplus}).  The target star is significantly 
smaller than the Sun ($\sim$0.74 R$_\odot$; B11), and the orbital periods of the candidates are shorter, so if confirmed the planets in this system 
would be the most closely spaced in terms of physical distance between orbits of any known system of several planets.  However, four of the planets 
orbiting KOI-500 appear to be locked in three-planet resonances (Subsection \ref{sec:res500}), whereas no analogous situation is observed in Kepler-11.  
Unfortunately, KOI-500 is quite faint, so it will be more difficult to study than is Kepler-11.  The four candidates in KOI-707 have periods within the 
range of the periods of the 5 inner planets observed in the Kepler-11 system. The multi-resonant KOI-730 (Subsection \ref{sec:res730}) is almost as closely-packed, but the resonances indicate a qualitatively 
different dynamical configuration.  Period ratios of neighboring candidates in the four planet KOI-117 system are all less than two, but nonetheless 
significantly larger than in Kepler-11 and KOI-707.  None of the five other four-candidate systems is nearly as closely spaced. Among the 45 
three-planet systems, only two, KOI-156 and KOI-168, have both neighboring pair period ratios less than 1.8.

Considering only super-Earth-size and Neptune-size planets with periods $< 125$ days, we conclude that Kepler-11 appears to be an extreme member of a 
class of very flat, closely-packed, planetary systems.  This class of systems seems to be significantly less common than the classes of planetary 
systems with a single planet and those that yield most of \ikt's multi-planet detections (Section \ref{sec:coplanar}), but nonetheless accounts for 
roughly 1\% of \ikt's targets that have planetary candidates. Kepler-11 is a relatively bright star ($Kp=13.7$), and two of the planets have relatively low SNR ($\sim$35), so it may be that similar systems that include somewhat smaller planets around fainter \ik targets will be revealed with additional data.

\section{Conclusions}

Analysis of the first four and one-half months of \ik data reveals 170 targets with more than one transiting planet candidate (B11).  While the vast 
majority of these candidates have yet to be validated or confirmed as true planets, we expect that the fidelity of this subsample of \ik planet candidates is 
high (Section \ref{sec:reli}), and the small fraction of false positives not to affect the robustness of our statistical results.  Incompleteness of 
the sample due to photometric noise and uncertainties in stellar parameters provide additional complications that could affect our results in a 
manner that is difficult to quantify. Many of our quantitative findings thus do not have error bars associated with them. Nonetheless, multi-transiting systems from \ik provide a large and rich dataset that can be used to powerfully test theoretical predictions 
of the formation and evolution of planetary systems.

Our major conclusions are:
\begin{enumerate}

\item The large number of candidate multiple transiting planet systems observed by \ik show that nearly coplanar multi-planet systems are 
common 
in short-period orbits around other stars.  This result holds for planets in the size range of $\sim$ 1.5 -- 6 R$_\oplus$, but not for giant planets 
(Figure \ref{fig:perrad}). 
Not enough data are yet available to assess its viability for Earth-size and smaller worlds, nor for planets with orbital periods longer than a few 
months.

\item Most  multiple planet candidates are neither in nor very near mean-motion orbital resonances.  Nonetheless, such resonances and near resonances are clearly more numerous than would be the case if period ratios were random.  First-order resonances dominate, but second-order resonances also are manifest. There appear to be at least three classes of resonance-related relationships evident in the data: The most abundant are planet pairs that have period ratios from one to a few percent larger than those of nearby resonances. Some planet pairs as well as chains of 
three or more planets have orbital periods within one part in 1000 of exact first-order resonance ratios. A few nearby pairs of planets 
deviate from exact period ratios by of order 1\% but appear to be protected from close approaches by resonantly librating 
configurations.  We note several 
systems that appear to have particularly interesting resonance configurations 
in Section \ref{sec:res}.  

\item Almost all candidate systems survived long-term dynamical integrations that assume circular, planar orbits and a mass-radius 
relationship (Equation \ref{eq:mr}) derived from planets within our Solar System (Section \ref{sec:stability}). This bolsters the evidence for the fidelity of the sample.

\item Taken together, the properties of the observed candidate multi-transiting systems suggest that the majority are free from false positives. Those 
candidates very near mean-motion resonances are most likely to be true planetary systems. Multi-transiting systems are extremely valuable for the study 
of planet 
formation, evolution, and dynamics \citep{Ragozzine:2010}.

\item Simulated ensemble populations of planetary systems generated to match the observed \ik multiple transiting planet systems tend to underpredict the observed number of singly transiting planets. This provides some evidence for a separate population or subpopulation of systems that either contains only one detectable planet per star or multiple 
planets with high relative orbital inclinations. A third, rarer, group of nearly co-planar densely-packed multi-planet systems (Section \ref{sec:K11}), also appears to be present. Note that these populations of planetary systems need not be cleanly separated, i.e., there may also be significant numbers of intermediate systems.

\item Approximately 3--5\% of \ik target stars have multiple planets in the 1.5 R$_\oplus < R_p < 6 $ R$_\oplus$ 
and $3 < P < 125$ day range (Section \ref{sec:coplanar}).

\item The inclination dispersion of most multiple planet systems in the above size-period range appears to have a mean of $\lesssim$10$^{\circ}$, suggesting relatively low mutual 
inclinations similar to the Solar System. Many singly-transiting systems may come from an additional population of systems with lower multiplicities and/or higher inclination dispersions. 

\item Many \ik targets with one or two observed transiting planet candidate(s) must be multi-planet systems where additional planets are present that are either not transiting and/or too small to be detected (Figure \ref{fig:simpopperradmore}). Some of known single candidates already show 
TTVs \citep{Ford:2011}, and we expect that many more will eventually be found to have such TTVs and that future \ik observations will reveal large numbers of new candidates in existing systems.

\end{enumerate}

The rich population of multi-transiting systems discovered by \ik and reported by B11 have immense value both as individuals and collectively for 
improving our understanding of the formation and evolution of planetary systems. The \ik spacecraft is scheduled to continue to return data on these 
multi-planet systems for the remainder of its mission, and the longer temporal baseline afforded by these data will allow for the discovery of more 
planets and more accurate measurements of the planets and their interactions. An extension of the mission well beyond the nominal 3.5 years would provide substantially better data for this class of studies.

\acknowledgments 
{\it Kepler} was competitively selected as NASA's tenth Discovery mission. Funding for this mission is 
provided by NASA's Science Mission Directorate. The authors thank the many people who gave so generously of their time 
to make the \ik mission a success. D. C. F. and J. A. C. acknowledge NASA support through Hubble Fellowship grants 
\#HF-51272.01-A and \#HF-51267.01-A, respectively, awarded by STScI, operated by AURA under contract NAS 5-26555. We 
thank Bill Cochran, Avi Loeb, Hanno Rein, Subo Dong, and Bill Welsh for valuable discussions and Kevin Zahnle, Tom Greene, Andrew Youdin, and an anonymous reviewer for constructive comments on the manuscript. Numerical integrations to test stability of nominal planetary systems were run on the supercomputer Pleiades at University of California, Santa 
Cruz.

\bibliographystyle{apj}

\end{document}